\documentclass[conference]{IEEEtran}
\IEEEoverridecommandlockouts

\usepackage{cite}
\makeatletter
\def\citepunct{,\penalty\@m\ } 
\makeatother

\usepackage{booktabs}
\usepackage{mathrsfs}
\usepackage{balance}
\DeclareMathAlphabet{\mathscr}{U}{rsfs}{m}{n}

\usepackage{amsthm}
\usepackage{amssymb}

\usepackage{xspace}
\usepackage{mathtools}
\usepackage{enumitem}
\usepackage{cleveref}
\usepackage[tikz]{mdframed}
\usepackage{framed}
\usepackage{subcaption}
\usepackage[ruled, vlined, linesnumbered]{algorithm2e}
\usepackage{makecell}
\usepackage{etoolbox}
\usepackage{multirow}

\usepackage[dvipsnames]{xcolor}
\newcommand{\revise}[1]{#1}

\newtoggle{comments}
\togglefalse{comments}

\iftoggle{comments}{
  \newcommand{\yw}[1]{{\texttt{\color{blue} Yue: [{#1}]}}}
  \newcommand{\jiongli}[1]{{\texttt{\color{magenta} Jiongli: [{#1}]}}}
}{
  \newcommand{\yw}[1]{}
  \newcommand{\jiongli}[1]{}
}

\newcommand{\baseline}{\texttt{PerColumn}\xspace}
\newcommand{\PerQuery}{\texttt{PerQuery}\xspace}

\newcommand{\sys}{\textsc{Mint}\xspace}
\newcommand{\speedupLowerBound}{2.1\times}
\newcommand{\speedupUpperBound}{8.3\times}

\newcommand{\SynNaive}{\texttt{Na\"ive}\xspace}
\newcommand{\SynLargeSimple}{\texttt{BiSimple}\xspace}
\newcommand{\SynLargeComplex}{\texttt{BiComplex}\xspace}
\newcommand{\News}{\texttt{News}\xspace}
\newcommand{\Amazon}{\texttt{Amazon}\xspace}

\newtheorem{thm}{Theorem}

\newtheorem{expl}{Example}
\newtheorem{dfn}{Definition}
\newtheorem{problem}{Problem}

\begin{document}
\setlength{\abovedisplayskip}{0pt}
\setlength{\belowdisplayskip}{0pt}
\setlength{\abovedisplayshortskip}{0pt}
\setlength{\belowdisplayshortskip}{0pt}

\title{MINT: Multi-Vector Search Index Tuning
}

\author{\IEEEauthorblockN{Jiongli Zhu\textsuperscript{*} \thanks{\textsuperscript{*}Work done at Microsoft Research.}} 
\IEEEauthorblockA{\textit{University of California, San Diego} \\
jiz143@ucsd.edu}
\\
\IEEEauthorblockN{Philip A. Bernstein}
\IEEEauthorblockA{\textit{Microsoft Research} \\
philbe@microsoft.com}
\and
\IEEEauthorblockN{Yue Wang}
\IEEEauthorblockA{\textit{Microsoft Research} \\
YueWang.UM@gmail.com}
\\
\IEEEauthorblockN{Vivek Narasayya}
\IEEEauthorblockA{\textit{Microsoft Research} \\
viveknar@microsoft.com}
\and
\IEEEauthorblockN{Bailu Ding}
\IEEEauthorblockA{\textit{Microsoft Research} \\
bailuding@gmail.com}
\\
\IEEEauthorblockN{Surajit Chaudhuri}
\IEEEauthorblockA{\textit{Microsoft Research} \\
surajitc@microsoft.com}
}

\maketitle

\begin{abstract}
Vector search plays a crucial role in many real-world applications. In addition to single-vector search, multi-vector search becomes important for multi-modal and multi-feature scenarios today. In a multi-vector database, each row is an item, each column represents a feature of items, and each cell is a high-dimensional vector. 
In multi-vector databases, the choice of indexes can significantly impact the performance of vector search.
Although index tuning for relational databases has been extensively studied, index tuning for multi-vector search remains unclear and challenging. In this paper, we define multi-vector search index tuning and propose a framework to solve it. Specifically, given a multi-vector search workload, we develop algorithms to find indexes that minimize latency and meet storage and recall constraints. Compared to the baseline, our techniques achieve a $\speedupLowerBound$ to $\speedupUpperBound$ speedup in latency.
\end{abstract}


\section{Introduction}

Vector similarity search has become increasingly important in real-world applications, such as recommendation systems, semantic search, and retrieval-augmented generation (RAG)~\cite{huang2020embedding,asai2023retrieval,jing2024large,pan:VLDBJ2024:survey}. However, due to the high dimensionality and large volume of data involved, performing exact similarity search is often computationally intractable. To overcome this challenge, efficient vector search indexes have been developed to approximate search results at high accuracy~\cite{malkov:TPAMI2018:hnsw, subramanya:NEURIPS2019:diskann,
Chen:Neurips2021:SPANN, 
liu2014sk, 
Douze:Arxiv2024:faiss}, known as Approximate Nearest Neighbor (ANN) search. 

Many existing approaches focus on single-domain data, overlooking that real-world datasets often contain information across multiple modalities, such as text, image, and video. For example, consider a product search on an e-commerce platform, where a user looks for ``100\% cotton T-shirt'' with a reference picture of a white graphic T-shirt attached. If the search system relies only on text, it may miss visual details like the white color or the graphic on the T-shirt, which are essential for capturing the user's preference for the T-shirt's style. On the other hand, relying solely on the image might overlook crucial details, such as ``100\% cotton'', that are not apparent from the image. By integrating both text and image data, a multi-modal search system can capture the user preference more holistically, leading to more accurate and relevant search results.
Recent research has shown that incorporating information from different modalities can significantly improve the accuracy of retrieval systems~\cite{
baltruvsaitis2018multimodal,
wang:ICDE2024:must,
duong2021efficient,
tautkute2017looks,ding2025www}.

There have been many studies of multi-vector search over multi-modal information~\cite{pan:VLDBJ2024:survey, wang:SIGMOD2021:milvus, zhang:OSDI2023:vbase, Chen:WWW2024:OneSparse, wang:ICDE2024:must}. 
Each item is represented by multiple vectors, each of which is an embedding from a modality or a feature. In a database of items, each row is one item, each column is one feature, and each cell is a high-dimensional vector. A ``multi-vector search'' query is represented by several vectors, each of which corresponds to a feature. The query can be on all columns (i.e., features) or a subset of columns. A score function, defined between a query and an item, is an aggregation of feature-level scores, which are based on vector similarity, such as dot product or cosine distance. The query result is a set of top-$k$ items with maximal scores.

\begin{sloppypar}
One approach to answer multi-vector queries is ``one-index-per-query''~\cite{pan:VLDBJ2024:survey, pan:SIGMOD2024:tutorial}.
For example, let $q^{1,3}$ be a query on column-$1$ and column-$3$. This approach creates an ANN index $x^{1,3}$ on both columns. 
Index $x^{1,3}$ answers $q^{1,3}$ efficiently, but cannot be used to fully answer
another query, say $q^{1,2,3}$ on column-$1$, column-$2$, and column-$3$. 
Therefore, one index must be created for each query, which consumes a lot of storage space and has long construction time. 
When the number of distinct column combinations on which queries are executed grows, this approach can 
become costly.
\end{sloppypar}

\begin{sloppypar}
The second approach is ``one-index-per-column''. It creates several single-column indexes and uses multiple indexes to answer each query \cite{wang:SIGMOD2021:milvus, zhang:OSDI2023:vbase}. In the above example of $q^{1,3}$ and $q^{1,2,3}$, one-index-per-column will create three indexes: $x^1$, $x^2$, and $x^3$. 
It answers $q^{1,3}$ by extracting items from $x^1$ and $x^3$ and then re-ranking all items together.
Similarly, it answers $q^{1,2,3}$ using $x^1$, $x^2$, and $x^3$ together. The number of indexes is limited by the number of columns, so the storage space is constrained. However, more work is needed to evaluate queries, since
the nearest neighbor of $q^{1,3}$ may not be a nearest neighbor in $x^1$ or $x^3$.
Thus many candidate neighbors must be extracted from $x^1$ and $x^3$ to achieve acceptable recall for $q^{1,3}$.
\end{sloppypar}

To overcome the limitations of the existing approaches, we propose 
answering queries using multi-column indexes.
Given a query, we develop a new query execution strategy that retrieves items from multi-column and/or single-column indexes. A query like $q^{1,2,3}$ may use an index created on a subset of the query's columns like $x^{1,3}$ and $x^2$. 
Multi-column indexes have better quality than single-column indexes, resulting in faster query evaluation than using the one-index-per-column approach.
And any query over a superset of the columns of a multi-column index will find the index to be useful.
So we avoid creating one index for each query. Our storage size is configurable and much smaller than one-index-per-query. 

Using this new query execution strategy, we develop a novel algorithm for choosing the set of indexes (called a configuration) and a novel query planner. 
Similar to the classical index tuning in relational databases~\cite{Chaudhuri:VLDB1997:AutoAdmin},
given a query workload and constraints for recall and storage, the configuration searcher explores the space of possible indexes to choose a configuration that minimizes the workload's latency while meeting the recall and storage constraints. 
The configuration searcher performs what-if query planning~\cite{Surajit:SIGMOD1998:WhatIf};
for each query, it asks for the best query plan for a given set of indexes. 
These indexes are hypothetical indexes based on sampling, which are much less expensive than creating real indexes. 
Like a relational database query optimizer, the query planner uses a cost model.
Given a query, it decides how best to use the hypothetical indexes to generate an efficient query execution plan.

In summary, given a set of multi-vector queries, we study how to choose which multi-column indexes to create
and how to use them to answer queries.
Our contributions are: (1) We define the Multi-vector Search Index Tuning problem and its building block, the Query Planning problem. We prove they are NP-hard and propose algorithms to solve them.
(2) To solve these problems, we propose simplified models to estimate query latency and recall.
(3) We implement \sys, a flexible index tuning framework for multi-vector search. 
It works with pluggable modules that estimate index performance. 
We demonstrate its flexibility using two types of vector index, HNSW~\cite{malkov:TPAMI2018:hnsw} and DiskANN~\cite{subramanya:NEURIPS2019:diskann}.
(4) We conduct experiments to measure the performance of \sys. \sys recommends indexes on millions of vectors within minutes, and it achieves a $\speedupLowerBound$ to $\speedupUpperBound$ speedup over the baseline.

This paper is organized as follows: We define the problem and provide an overview in Section~\ref{section:Overview}. We present the two major modules, Query Planner and Configuration Searcher, in Section~\ref{section:QueryPlanner} and Section~\ref{section:ConfigurationSearcher}. We evaluate \sys in Section~\ref{section:Experiments}. Finally, we study related work in Section~\ref{section:RelatedWork} and conclude in Section~\ref{section:Conclusion}.


\section{Problem Definition and Overview}
\label{section:Overview}

In this section, we present the Multi-Vector Search Query and the Multi-Vector Search Index Tuning problem, challenges in solving the problem, and an overview of our solution.
We define the problem following the literature~\cite{pan:VLDBJ2024:survey, pan:SIGMOD2024:tutorial, wang:SIGMOD2021:milvus, zhang:OSDI2023:vbase}.

\begin{dfn}[Multi-Vector Search Query]
\label{definition:MultiVectorSearch}
Given a database of items, each item (i.e., row) $r$ is represented as $m$ vectors $\textbf{v}_1, \textbf{v}_2, ..., \textbf{v}_m$ (i.e., columns). A query $q$ is on a subset of vectors whose column-IDs are in $q.vid$. The query $q$ finds top-$k$ items that maximize $\sum_{i \in q.vid} score(q.\textbf{v}_i, r.\textbf{v}_i)$ where $score(\cdot, \cdot)$ is a score function.
\end{dfn}

In practice, the score function is usually based on dot product, cosine similarity, Hamming or Euclidean distance, or $L_p$-norm.

\begin{expl}
\revise{
Consider a database of items (e.g., e-commerce products) that represents each item using three vectors (e.g., product title, description, and main image). 
A query $q$ is on the 1st and 3rd vectors (i.e., $q.vid=\{1, 3\}$ on product title and main image). 
If the score function $score(\cdot,\cdot)$ is dot product, 
then query $q$ finds the top-$k$ items that maximize $\sum_{i \in q.vid} score(q.\textbf{v}_i, r.\textbf{v}_i) = q.\textbf{v}_1 \cdot r.\textbf{v}_1 + q.\textbf{v}_3 \cdot r.\textbf{v}_3$.
}
\end{expl}

Instead of brute-force scanning of all vectors, we can create indexes to answer a query more efficiently. For instance, HNSW~\cite{malkov:TPAMI2018:hnsw} 
indexes can be created on the first ($\textbf{v}_1$) and the third ($\textbf{v}_3$) column. 
Then we can retrieve top results from each index based on partial score ($q.\textbf{v}_1 \cdot r.\textbf{v}_1$) and ($q.\textbf{v}_3 \cdot r.\textbf{v}_3$), re-rank them using full score ($q.\textbf{v}_1 \cdot r.\textbf{v}_1 + q.\textbf{v}_3 \cdot r.\textbf{v}_3$), and return the top-$k$ results to the user~\cite{wang:SIGMOD2021:milvus, zhang:OSDI2023:vbase}.
Or we can create an index on the concatenation of the first and third columns, and directly retrieve top-$k$ results. 
Different choices of indexes can lead to different storage footprints 
and different query execution plans, with different latency and recall.

We now define the Multi-Vector Search Index Tuning problem:

\begin{problem}[Multi-Vector Search Index Tuning]
\label{problem:MultiVectorIndexTuning}
Given a database of items and a query workload $W = \{(q_1, p_1), (q_2, p_2), \allowbreak (q_3, p_3), ...\}$ where $q_i$ is a query and $p_i$ is the probability (i.e., normalized frequency) of getting $q_i$, find a set of indexes $X$ to answer the queries with lowest latency meeting recall and storage constraints:
\begin{flalign}
\min_{X}\quad & \sum_{(q_i, p_i) \in W} p_i \cdot latency(q_i, X) \label{formula:optObjective} \\
s. t.\quad & \forall (q_i, p_i) \in W, recall(q_i, X) \geq \theta_{recall} \label{formula:optRecall} \\
& storage(X) \leq \theta_{storage} \label{formula:optStorage}
\end{flalign}
where $latency(q_i, X)$ and $recall(q_i, X)$ are the latency and recall of answering $q_i$ using $X$, $storage(X)$ is the storage size of $X$, and $\theta_{recall}$ and $\theta_{storage}$ are the thresholds of recall and storage.
\end{problem}

In the above definition, the objective is to minimize the weighted latency of the workload (Formula~\ref{formula:optObjective}), subject to constraints on the recall (Formula~\ref{formula:optRecall}) and storage (Formula~\ref{formula:optStorage}). We use a single $\theta_{recall}$ for all queries for simplicity, but in practice, users can specify different recall thresholds for different queries when necessary.

We use superscripts to indicate the columns of a query or index. For example, $q^{1,3}$ is a query on the 1st and the 3rd columns, and $x^{1,3}$ is an index on the 1st and the 3rd columns. 
We use the term \textbf{configuration} to mean a set of indexes~\cite{chaudhuri:VLDB2007:TuningDecade}. 

\begin{expl}
\revise{
Assume a workload $W=\{ (q^{1,3}, 0.35), \allowbreak (q^{2}, 0.35), \allowbreak (q^{1,2,3}, 0.3)\}$ where $q^{1,3}$ is on the 1st and 3rd columns (e.g., product title and main image), $q^{2}$ is on the 2nd column (e.g., description), and $q^{1,2,3}$ is on all three columns. If we are allowed to create only two indexes, a possible configuration
is  $X=\{x^{1,3}, x^2\}$, where $x^{1,3}$ may answer $q^{1,3}$, $x^{2}$ may answer $q^2$, and $x^{1,3}$ together with $x^{2}$ may answer $q^{1,2,3}$.
}
\end{expl}

\subsection{Challenges}

To solve Problem~\ref{problem:MultiVectorIndexTuning}, we must address three major challenges:

\begin{enumerate}[noitemsep,nolistsep,leftmargin=*] 

\item \textbf{Finding the best query plan using given indexes}. 
When we explore the space of indexes and find a set of indexes for a given query, which subset of indexes should be used to answer the query to minimize the latency while meeting the recall threshold? 
We prove this problem is NP-hard in Section~\ref{section:QueryPlanningNP}.
Since trying all possible subsets is too expensive, we need an algorithm to quickly find a good plan for using indexes. 
We implement a query planner, whose role is conceptually similar to a query optimizer in relational databases but uses a completely different algorithm (Section~\ref{section:QueryPlanningAlgorithms}). The query planner takes a query and a configuration as input, and outputs a query plan with the estimated latency and recall. 

\item \textbf{Estimating latency and recall using simplified models}. The optimization problem in Problem~\ref{problem:MultiVectorIndexTuning} requires good estimates of latency (Formula~\ref{formula:optObjective}) and recall (Formula~\ref{formula:optRecall}) for each query. 
Latency has a large variance across different memory, CPU, GPU, etc. Therefore, we define \textbf{cost} as a proxy for latency. Cost is based on the number of distance computations~\cite{pan:VLDBJ2024:survey, malkov:TPAMI2018:hnsw, wei:VLDB2020:analyticDBV, zhang:OSDI2023:vbase}.

\begin{sloppypar}    
\hspace{1em} 
Many popular indexes, such as  HNSW~\cite{malkov:TPAMI2018:hnsw} and DiskANN~\cite{subramanya:NEURIPS2019:diskann}, do not have a closed-form solution for their costs.
Thus, index creation is probably unavoidable for estimating the cost of answering queries.
Na\"ively creating indexes on all items for all plans will take too long and consume too much space. 
For example, most graph-based indexes need hundreds of seconds to create for one million items~\cite{wang:VLDB2021:graphExperimentalComparison}. 
Even worse, the number of indexes can be exponential in the number of columns and number of plans to consider. 
To solve these issues, we borrow the idea of sampling items~\cite{Chaudhuri:Sigmod1998:Sampling} to limit the creation time of each index, and use single-column indexes to estimate the cost of multi-column indexes without creating them. 
\end{sloppypar}

\hspace{1em} Similarly, we apply these techniques to estimate the \emph{recall}, which also has no closed-form solution for many index types, including HNSW and DiskANN. 
More detail is in Section~\ref{section:RecallAndCostEstimator}. 

\hspace{1em} In theory, one can always construct an adversarial query requiring a linear scan of $O(n)$ vectors (i.e., expensive cost) to achieve an acceptable recall, where $n$ is the total number of rows~\cite{Indyk:Neurips2023:DiskANNWorstCase}, but in practice, our estimations work reasonably well as we show in experiments (Section~\ref{section:Experiments}).


\item \textbf{Finding the best configuration}. 
After we find the best query plan and estimate latency and recall for each individual query, the remaining challenge is to find the best configuration for the entire workload. 
Ideally, we would like to include the best indexes for all queries. 
However, this ideal configuration usually exceeds the storage threshold (Formula~\ref{formula:optStorage}). 
Therefore, we must pick a smaller set of indexes that meets the recall and storage thresholds while minimizing cost (i.e., latency). 
This problem proves to be NP-hard as we will see in Section~\ref{section:ConfigurationSearchNP}. So in Section~\ref{section:ConfigurationSearchAlgorithm}, we develop a search algorithm that finds configurations.
It outperforms the baseline by $\speedupLowerBound$ to $\speedupUpperBound$.

\end{enumerate}

\begin{figure}[t]
    \centering
    \includegraphics[width=.8\linewidth]{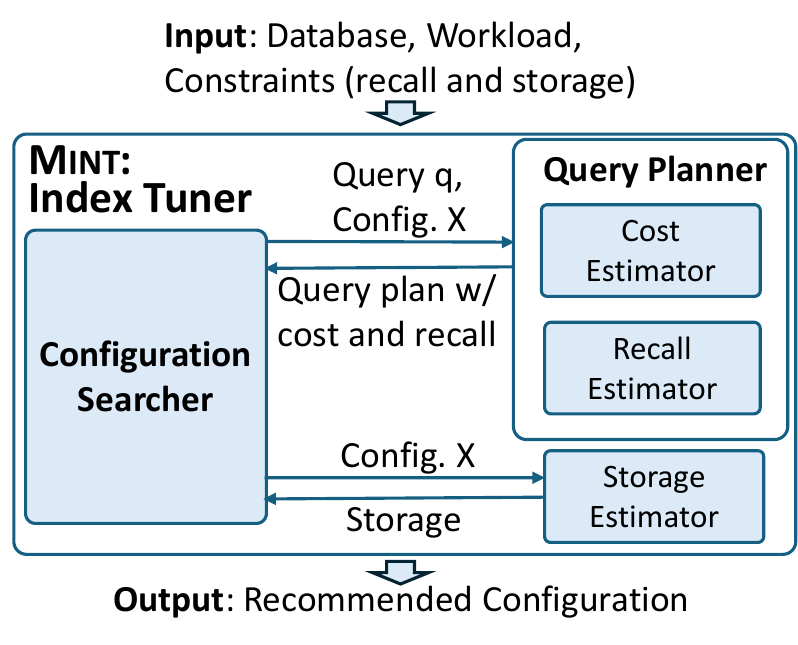}
    \vspace{-3mm}
    \caption{Overview of \sys. The Configuration Searcher searches for the workload's best configuration by interacting with the Query Planner and the Storage Estimator. The Query Planner finds each individual query's best plan using the Cost Estimator and the Recall Estimator.}
    \label{fig:overview}
\end{figure}

\subsection{Overview}
\label{section:FrameworkOverview}

Figure~\ref{fig:overview} presents an overview of the major modules that address the above challenges. Our Index Tuner, \sys, takes the database, workload, and constraints as input and returns a recommended configuration as output. Within the tuner, the Configuration Searcher searches for the workload's best configuration by interacting with the Query Planner and Storage Estimator. The Query Planner takes a query and a configuration as input, and finds the best query plan using the Cost Estimator and Recall Estimator. The Storage Estimator is independent of the query. It only takes the configuration as input and outputs estimated storage.

\sys works with different indexes and estimators. We implemented sample-based estimators for HNSW and DiskANN.
Other estimators can be plugged in without changing the framework.

\begin{sloppypar}
The Storage Estimator in Figure~\ref{fig:overview} is straightforward for many types of index, including partition-based, tree-based, and graph-based. 
For instance, in graph-based indexes like HNSW and DiskANN, each item is a node with a fixed maximum {\em degree}, i.e., number of edges. 
In this case, the index's storage usage is the number of items times the degree and the size of each edge. \emph{Note} that each index only stores the graph using item IDs without duplicating the underlying vectors. So the index graph size is independent of the dimensions of vectors.\footnote{\revise{For some index techniques like LSH (which needs more than linear space~\cite{Mccauley:ESA2024:Space, Andoni:CommACM2008:LshAnn}) and subspace collision framework~\cite{Wei:SIGMOD2025:SubspaceCollision}, the index storage is not necessarily linear in the number of items. In such cases, technique-specific storage estimators are reqruied and can be plugged into our framework.}} 
Next, we will focus on Query Planner and Configuration Searcher.
\end{sloppypar}



\section{Query Planner}
\label{section:QueryPlanner}

Next, we define the Query Planning problem (Section~\ref{sec:DefQueryPlanning}), prove it is NP-hard (Section~\ref{section:QueryPlanningNP}), and develop Dynamic Programming algorithms to solve it (Section~\ref{section:QueryPlanningAlgorithms}).

\subsection{Definition of Query Planning} 
\label{sec:DefQueryPlanning}

Given a query and a configuration (i.e., a set of indexes), Query Planning involves finding a 
query plan that uses a subset of indexes and minimizes cost (i.e., latency) while meeting the recall requirement. 


The cost comes from two sources, index scan and re-ranking, and depends on how many items we retrieve from each index. 
Consider the following example to understand what a query plan looks like and what cost it incurs.

\begin{figure}
    \centering
    \includegraphics[width=.8\linewidth]{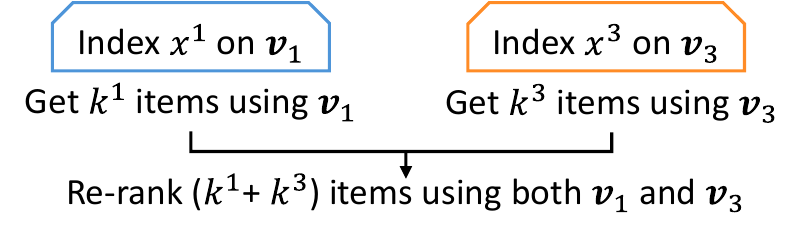}
    \vspace{-3mm}
    \caption{An example query plan retrieves $k^1$ and $k^3$ items from $x^1$ and $x^3$ and then re-rank.}
    \label{fig:QueryPlan}
\end{figure}

\begin{expl}
\label{example:QueryPlan}
A query $q^{1,3}$ gets the top-$k$ items based on the first and third columns  ($\textbf{v}_1$ and $\textbf{v}_3$).
Suppose a query plan (Figure~\ref{fig:QueryPlan}) uses indexes $x^1$ on $\textbf{v}_1$ and $x^3$  on $\textbf{v}_3$ to answer $q^{1,3}$.
The plan 
has three steps.
\begin{enumerate}[noitemsep,nolistsep,leftmargin=1.5em]
\begin{sloppypar}
\item  Use index $x^1$ to retrieve $k^1$ items using $score(q^{1,3}.\textbf{v}_1, r.\textbf{v}_1)$ where $r$ is the item in the database. 
\item Use index $x^3$ to retrieve $k^3$ items using $score(q^{1,3}.\textbf{v}_3, r.\textbf{v}_3)$. 
\item Re-rank the $(k^1+k^3)$ items using
$score(q^{1,3}.\textbf{v}_1, r.\textbf{v}_1) + score(q^{1,3}.\textbf{v}_3, r.\textbf{v}_3)$
to return the top-$k$ items as the result. 
\end{sloppypar}
\end{enumerate}
\end{expl}
$k^1$ and $k^3$ are 
parameters of the query plan.
Different parameters lead to different plans with different costs. 
We refer to $k^1$ and $k^3$ as ``extended-$k$'', denoted $ek$, for individual index scans to distinguish them from the given $k$. 

A query plan can be viewed as $(X, EK)$ where $X=\{x_i\}$ is the indexes and $EK=\{ek_i\}$ is the number of items retrieved. 
It means that the top-$ek_i$ items are retrieved from index $x_i$ and finally re-ranked. 
Intuitively, a greater $ek_i$ means more items are retrieved, yielding a higher cost and higher recall, while $ek_i=0$ means index $x_i$ is not used and incurs no cost.

Formally, we define cost as a function of the query $q$, configuration $X$, and index scan parameters $EK$:
{\footnotesize
\begin{align} \label{eq:CostConf} 
cost_{plan}(q, X, EK) = \left( \sum_{i = 1}^{|X|} cost_{idx}(q, x_i, ek_i) \right) + cost_{rerank}(q, EK)
\end{align}
}
where $cost_{idx}(q, x_i, ek_i)$ is the cost of retrieving the top $ek_i$ items from the index scan of $x_i$, and $cost_{rerank}(q, EK)$ is the cost of re-ranking.

For each individual index scan, we define the cost of retrieving top-$ek_i$ items from $x_i$ as: 
\begin{align}
\label{equation:CostIndexScan}
cost_{idx}(q, x_i, ek_i) = x_i.dim \cdot numDist(q, x_i, ek_i)
\end{align}
where $x_i.dim$ is the dimension of $x_i$ and $numDist(q, x_i, ek_i)$ is the number of distance computations (i.e., number of items compared). We take $x_i.dim$ into consideration because the dimensions differ across vector embeddings. For instance, computing a distance on a $100$-dimensional index is much faster than on a $200$-dimensional index. 
$numDist(q, x_i, ek_i)$ is simply the number of invocations of the score function during the index scan.

The re-ranking cost is captured in $cost_{rerank}(q, EK)$. For simplicity, we consider only the dominating cost of high-dimensional distance computation and omit the cost of getting top $k$ scalars using a priority queue.
Ideally, the cost should be based on the number of distinct item IDs because duplicate items are computed only once. But it is hard to estimate distinct item IDs. Therefore, we simplify it by including duplicate IDs and counting only $ek_i$:
\begin{align}
cost_{rerank}(q, EK)=q.dim \cdot \sum_{i=1}^{|EK|} ek_i    
\end{align}
where $q.dim$ is the dimension of $q$.

Next, we define recall by comparing the retrieved items with the ground truth. Let the ground truth item IDs be $gt(q)$. 
Then let $res(q, x_i, ek_i)$ be the $ek_i$ item IDs retrieved from index $x_i$. The recall is defined on the union of all retrieved item IDs:
\begin{align} \label{eq:RecallQInst}
recall(q, X, EK) = \frac{\left|gt(q) \cap \left(\cup_{i=1}^{|X|} res(q, x_i, ek_i)\right) \right|}{|gt(q)|}
\end{align}

Now we can define a query planning problem:
\begin{problem}[Query Planning]
\label{problem:SimplifiedQO}
Given a query $q$ and a configuration $X$, find $EK$ that minimizes the cost while satisfying the recall threshold $\theta_{recall}$.
\begin{flalign*}
\min_{EK}\quad & cost_{plan}(q, X, EK)  \\
s. t.\quad & recall(q, X, EK) \geq \theta_{recall}
\end{flalign*}
\end{problem}






\subsection{NP-Hardness of Query Planning}
\label{section:QueryPlanningNP}

We can prove that Problem~\ref{problem:SimplifiedQO} is NP-hard by reducing the Set Cover problem to it.
The Set Cover problem is to find a set of subsets to cover the universe set. Formally,
\begin{problem}[Set Cover $(\mathtt{U}, \mathtt{S}, \mathtt{t})$]
\label{problem:SetCover}
Given a universe set $\mathtt{U}=\{1, 2, 3, ..., \mathtt{n}\}$ and a set of subsets $\mathtt{S}=\{\mathtt{s}_i\}$ each of whose elements $\mathtt{s}_i$ is a subset of $\mathtt{U}$, a set $\mathtt{C} \subseteq \mathtt{S}$ is a set cover if the union of elements of $\mathtt{C}$ equals $\mathtt{U}$. Find a set cover $\mathtt{C}$ such that $|\mathtt{C}| \leq \mathtt{t}$.
\end{problem}
\begin{thm}
Problem~\ref{problem:SimplifiedQO} is NP-hard.
\end{thm}

Intuition: We reduce the Set Cover problem to Problem~\ref{problem:SimplifiedQO} by making the universe set $\mathtt{U}$ be ground truth, and making each index $x_i$ ``return $\mathtt{s}_i$ in limited cost'' if and only if $ek_i = |\mathtt{U}| = \mathtt{n}$. More specifically, when $ek_i < \mathtt{n}$, design $res(q,x_i,ek_i)$ to make index $x_i$ return nothing in $\mathtt{s}_i$; when $ek_i > \mathtt{n}$, design $cost_{idx}(q,x_i,ek_i)$ to make the cost huge.

\begin{proof} \textbf{(Sketch)}
We first reduce an arbitrary Set Cover problem to Problem~\ref{problem:SimplifiedQO} in polynomial time. Given $(\mathtt{U}, \mathtt{S}, \mathtt{t})$ in Set Cover, let $gt(q)=\mathtt{U}$ and 
$q.dim = |\mathtt{S}|=\mathtt{d}$, map each $\mathtt{s}_i \in \mathtt{S}$ to an index $x_i$ with $x_i.dim=1$, and define
\begin{align*}
cost_{idx}(q, x_i, ek_i) = 
\left\{
\begin{aligned}
0 &\quad \text{when}~ek_i=0, \\
\mathtt{n} &\quad \text{when}~1\leq ek_i \leq \mathtt{n}-1, \\
2\mathtt{n}+1 &\quad \text{when}~ek_i=\mathtt{n}, \\
\infty &\quad \text{when}~ek_i\geq\mathtt{n}+1.
\end{aligned}
\right. 
\end{align*}

$res(q, x_i, ek_i) =$
\begin{align*}
\left\{
\begin{aligned} 
\small
\text{a set of $ek_i$ dummy IDs}  &\quad \text{when}~0\leq ek_i \leq \mathtt{n}-1, \\
\small
\mathtt{s}_i \text{~and~} (ek_i-|\mathtt{s}_i|) \text{~dummy IDs} &\quad \text{when}~ek_i\geq \mathtt{n}.
\end{aligned}
\right. 
\end{align*}

Such $res(q, x_i, ek_i)$ is possible in practice because in a graph-based index like HNSW, increasing $ek_i$ by 1 can lead to exploration of more than 1 items in HNSW and replace all top-$ek_i$ results. We will show an example of such an $x_i$ after the proof.

\begin{sloppypar}
Next, we map each Set Cover solution $\mathtt{C}$ to a Query Planning solution $EK$ and vice versa. Formally,
we show that a Set Cover problem $(\mathtt{U}, \mathtt{S}, \mathtt{t})$ has a solution $\mathtt{C}$ if and only if Problem~\ref{problem:SimplifiedQO} has a solution $EK$ satisfying $recall(q, X, EK)=100\%$ and $cost_{plan}(q, X, EK) \leq (2\mathtt{n}+1) \cdot \mathtt{t} + \mathtt{d} \mathtt{n} \cdot \mathtt{t}$.
\end{sloppypar}

\begin{sloppypar}
(A) We first prove the mapping from $\mathtt{C}$ to $EK$. 

Formally, given a Set Cover solution $\mathtt{C}$ where $|\mathtt{C}| \leq \mathtt{t}$, a solution $EK$ exists for Problem~\ref{problem:SimplifiedQO} satisfying $recall(q, X, EK)=100\%$ and $cost_{plan}(q, X, EK) \leq (2\mathtt{n}+1) \cdot \mathtt{t} + \mathtt{d} \mathtt{n} \cdot \mathtt{t}$.
\end{sloppypar}

For each $\mathtt{s}_i \in \mathtt{S}$, let $ek_i=\mathtt{n}$ if $\mathtt{s}_i \in \mathtt{C}$, otherwise $ek_i=0$.
\begin{itemize}[noitemsep,nolistsep]
\item $recall(q, X, EK)=100\%$ because $\mathtt{C}$ is a set cover that makes $\cup_{i=1}^{|X|} res(q, x_i, ek_i) = \cup_{\mathtt{s}_i \in \mathtt{C}} (\mathtt{s}_i \text{~and dummy IDs}) \supseteq \mathtt{U}$.

\item $cost_{plan}(q, X, EK) \leq (2\mathtt{n} + 1) \cdot \mathtt{t} + \mathtt{d} \mathtt{n} \cdot \mathtt{t}$ because by Equation~\ref{eq:CostConf}:
\begin{align*}
& cost_{plan}(q, X, EK) \\
=&\quad \textstyle \sum_{i = 1}^{|X|} cost_{idx}(q, x_i, ek_i) + cost_{rerank}(q, EK) \\
=&\quad \textstyle \sum_{i = 1}^{|X|} \Big(cost_{idx}(q, x_i, ek_i) + \mathtt{d} \cdot ek_i \Big)
\end{align*}
For those $ek_i=0$, the cost is 0. For those $ek_i=n$, the cost is $(2\mathtt{n} + 1 + \mathtt{d} \mathtt{n})$.
So $cost_{plan}(q, X, EK) = |\mathtt{C}| \cdot (2\mathtt{n} + 1 + \mathtt{d} \mathtt{n}) \leq \mathtt{t} \cdot (2\mathtt{n} + 1 +  \mathtt{d} \mathtt{n})$ 
\end{itemize}

\begin{sloppypar}
(B) We now prove the mapping from $EK$ to $\mathtt{C}$.

Formally, given a solution $EK$ satisfying $recall(q, X, EK)=100\%$ and $cost_{plan}(q, X, EK) \leq (2\mathtt{n} + 1) \cdot \mathtt{t} + \mathtt{d} \mathtt{n} \cdot \mathtt{t}$, a set cover $\mathtt{C}$ exists and $|\mathtt{C}| \leq \mathtt{t}$.
\end{sloppypar}

Let $\mathtt{C}=\{\mathtt{s}_i ~|~ ek_i \geq \mathtt{n}\}$. 
\begin{itemize}[noitemsep,nolistsep]
\item We prove $\mathtt{C}$ is a set cover:
$recall(q, X, EK)=100\%$ means $\cup_{\mathtt{s}_i \in \mathtt{C}} = \mathtt{U}$ because dummy IDs do not impact recall.

\item Prove $|\mathtt{C}| \leq \mathtt{t}$:

\begin{sloppypar}
By Equation~\ref{eq:CostConf}, $cost_{plan}(q, X, EK) = \sum_{i = 1}^{|X|} cost_{idx}(q, x_i, ek_i) \allowbreak + \allowbreak cost_{rerank}(q, EK) = \sum_{i = 1}^{|X|} \Big(cost_{idx}(q, x_i, ek_i) + \mathtt{d} \cdot ek_i \Big)$.
\end{sloppypar}

Since $cost_{plan}(q, X, EK) \leq (2\mathtt{n}+1) \cdot \mathtt{t} + \mathtt{d} \mathtt{n} \cdot \mathtt{t}$, there must be \emph{no} $ek_i \geq \mathtt{n}+1$ that makes $cost_{idx}(q, x_i, ek_i)=\infty$. 

$\therefore$ $cost_{plan}(q, X, EK) = \sum_{1 \leq ek_i \leq \mathtt{n} - 1} (\mathtt{n} + \mathtt{d} \cdot ek_i) + \sum_{ek_i = \mathtt{n}}(2\mathtt{n}+1 + \mathtt{d} \cdot ek_i) = \sum_{1 \leq ek_i \leq \mathtt{n} - 1} (\mathtt{n} + \mathtt{d} \cdot ek_i) + |\mathtt{C}|(2\mathtt{n}+1 + \mathtt{d} \cdot \mathtt{n})$.


Because $\sum_{1 \leq ek_i \leq \mathtt{n} - 1} (\mathtt{n} + \mathtt{d} \cdot ek_i) \geq 0$, 

$\therefore |\mathtt{C}|(2\mathtt{n}+1 + \mathtt{d} \cdot \mathtt{n}) \leq cost_{plan}(q, X, EK) \leq (2\mathtt{n}+1) \cdot \mathtt{t} + \mathtt{d} \mathtt{n} \cdot \mathtt{t}$


$\therefore |\mathtt{C}| \leq \mathtt{t}$.
\end{itemize}

\end{proof}

\begin{sloppypar}
We provide an example in which the designed $cost_{idx}(q,x_i,ek_i)$ and $res(q,x_i,ek_i)$ are real. The example is an HNSW index with one layer graph. Given $ek_i$, the HNSW search algorithm~\cite{malkov:TPAMI2018:hnsw} starts from a fixed entry point and explores neighbors to update the top-$ek_i$ items. The search stops when the new neighbors cannot make top-$ek_i$ items better (i.e., top-$ek_i$ items converge). That means, even when we increase $ek_i$ by $1$, more than one neighbor can be explored. As we will see, when $ek_i \leq \mathtt{n}-1$, the search stops after scanning $\mathtt{n}$ items. When $ek_i = \mathtt{n}$, the search stops after scanning $(2\mathtt{n}+1)$ items. When $ek_i \geq \mathtt{n}+1$, the search has to scan a large number of items (i.e., all items in the database).
\end{sloppypar}

\begin{figure}
    \centering
    \includegraphics[width=.68\linewidth]{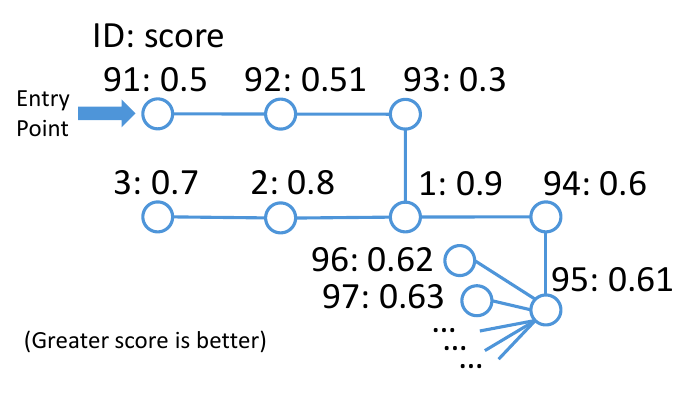}
    \caption{Example HNSW index with one layer for $\mathtt{n}=3$. IDs $\{1, 2, 3\} = \mathtt{U}$ in Problem~\ref{problem:SetCover}, while $\text{IDs}\geq91$ are dummy IDs. When $1 \leq ek_i\leq \mathtt{n}-1 = 2$, $\mathtt{n}=3$ items in the first row are scanned. When $ek_i=\mathtt{n}=3$, $(2\mathtt{n}+1)=7$ items in the first and seconds rows are scanned. When $ek_i \geq \mathtt{n}+1$ = 4, all items in the database are scanned.}
    \label{fig:proofExampleHnsw}
\end{figure}

\begin{expl}
Figure~\ref{fig:proofExampleHnsw} is an HNSW index with one layer graph. Let $\mathtt{n}=3$. IDs $1$ to $3$ are in the universe set $\mathtt{U}$ in Problem~\ref{problem:SetCover}, while $\text{IDs}\geq91$ are dummy IDs. 
\begin{itemize}[noitemsep, nolistsep,leftmargin=2em]
\item When $ek_i = 0$, the index is not scanned. $cost_{idx}(q,x_i,ek_i) = 0$ and $res(q,x_i,ek_i) = \varnothing$.

\item When $1 \leq ek_i \leq \mathtt{n} - 1 = 2$, the search explores items $91$ and $92$ and stops because $92$'s new neighbor $93$ does not improve the top items. $cost_{idx}(q,x_i,ek_i) = \mathtt{n} = 3$ because the scores of $\{91, 92, 93\}$ are computed. $res(q,x_i,ek_i)$ contains only dummy IDs.

\item When $ek_i = \mathtt{n} = 3$, the search further explores $93$'s neighbor to find better items $\{1, 2, 3\}$, and stops because item $1$'s neighbor $94$ does not improve top-3. So $cost_{idx}(q,x_i,ek_i) = 2\mathtt{n}+1 = 7$ for $\{1,2,3,91,92,93,94\}$. $res(q,x_i,ek_i)=\{1,2,3\}$.

\item When $ek_i \geq \mathtt{n} + 1 = 4$, the search further explores $94$'s neighbor to find $95$, and then has to explore all remaining items like $96$, $97$, etc. So $cost_{idx}(q,x_i,ek_i) \approx \infty$. $res(q,x_i,ek_i)=\{1,2,3\} \cup \{\text{dummy IDs}\}$.
\end{itemize}
\end{expl}

\subsection{Dynamic Programming Solution for Query Planning}
\label{section:QueryPlanningAlgorithms}

\subsubsection{Search and Dynamic Programming Algorithms}~

Although Problem~\ref{problem:SimplifiedQO} is NP-hard, we can address it by leveraging the limited $k$ and $EK$ in practice. In real-world applications, $k$ is usually around $10$ to a few hundred~\cite{Aumuller:InfoSys2020:ANN-Benchmarks, Simhadri:Neurips2021:BigANN-Benchmark, Simhadri:Arxiv2024:BigANN-Benchmark}. 
Assume for a moment that we can get the ground truth $gt(q)$ and $cost_{idx}(q, x_i, ek_i)$; we will relax this assumption and discuss how to estimate them in Section~\ref{section:RecallAndCostEstimator}.
For each item in the ground truth $gt(q)$, we can identify the item's ranking in each index $x_i$ as a \emph{relevant} $ek_i$. In other words, for $x_i$, the number of relevant $ek_i$ is limited to $k$.

\begin{figure}
    \centering
    \includegraphics[width=.83\linewidth]{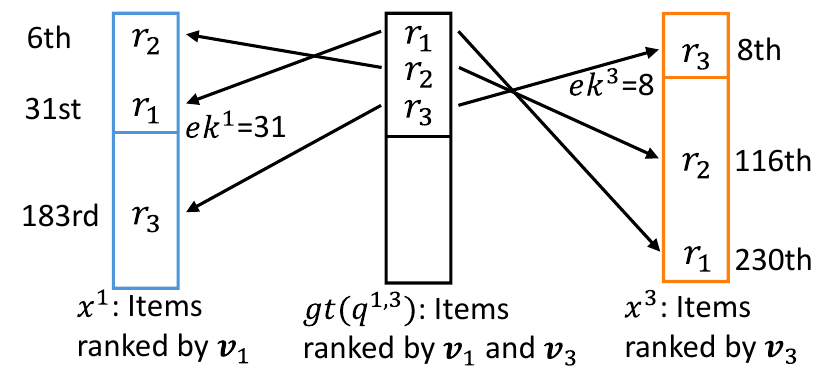}
    \caption{Example of relevant $ek$. The number of each relevant $ek$ is limited to $k=3$. The search space of $ek^1$ and $ek^3$ is $\{0, 6, 31, 183\} \times \{0, 8, 116, 230\}$, which has $(k+1)(k+1)=16$ combinations. A possible plan is $ek^1=31$ and $ek^3=8$, which covers $r_1$, $r_2$, and $r_3$.}
    \label{fig:relevantK}
\end{figure}

\begin{expl}
(Continuing Example~\ref{example:QueryPlan}) 
Let $k=3$ for query $q^{1,3}$ and the indexes be $x^1$ on $\textbf{v}_1$ and $x^3$ on $\textbf{v}_3$. 

Figure~\ref{fig:relevantK} illustrates the search space. Instead of exploring all possible $ek^1 \geq 0$ for $x^1$ and $ek^3 \geq 0$ for $x^3$, we only focus on $k$ relevant $ek^1$ and $k$ relevant $ek^3$.

Take $ek^1$ as an example. For the $j$-th item in the ground truth $gt(q^{1,3})$ ranked by $\sum_{i \in \{1,3\}} score(q^{1,3}.\textbf{v}_i, r.\textbf{v}_i)$,
we can find its ranking $ek^1.j$ in $x^1$ by $score(q^{1,3}.\textbf{v}_1, r.\textbf{v}_1)$. 
For instance, the 1st item in $gt(q^{1,3})$ may have $ek^1.1=31$ in $x^1$, the 2nd item in $gt(q^{1,3})$ may have $ek^1.2=6$, and so on. If no item in $gt(q^{1,3})$ ranks between $7$ and $30$, we do not care $7 \leq ek^1 \leq 30$ because, compared to $ek^1=6$, $ek^1$ between $7$ and $30$ do not cover more items in $gt(q^{1,3})$. In other words, only the $3$ $ek^1.j$ are relevant.

\begin{sloppypar}
Similarly, we can find the $3$ rankings $ek^3.j$ in $x^3$ by $score(q^{1,3}.\textbf{v}_3, r.\textbf{v}_3)$.    
\end{sloppypar}

Then, we add $ek^1.0=0$ and $ek^3.0=0$ indicating the corresponding index may be skipped.

Finally, we only focus on search space $\{ek^1.j\} \times \{ek^3.j\}$ whose size is $(k+1) \cdot (k+1)$.
\end{expl}


Because of the limited number of relevant $ek_i$, we can enumerate the combinations of $ek_i$ to find the best plan (detail in~\cite{Zhu:Arxiv2025:Mint}). Given indexes $X$, the search space of $EK$ is $(k+1)^{|X|}$. The search algorithm's space complexity is $O(k \cdot |X|)$ for pre-computing the relevant $ek_i$. The time complexity is $O(k \cdot \sum_{x_i \in X} x_i.dim + |X| k \log k + k^{|X|})$ where $k \cdot \sum_{x_i \in X} x_i.dim$ is to compute the scores on $x_i$,  $|X| k \log k$ is to sort $ek_i.j$ for each $x_i$, and $k^{|X|}$ is to enumerate the combinations of $ek_i$ and incrementally counting the covered items.
The $k^{|X|}$ term limits its scalability. So we further optimize it by proposing a Dynamic Programming algorithm without $|X|$ as an exponent.

\begin{algorithm}[t]
\footnotesize
\KwIn{Query $q$; Top $k'$; Candidate indexes $X$; Recall threshold $\theta_{recall}$; Ground truth sample $gt_{sp}(q)$; Cost function $cost_{cover}(cvr, x_i)$.}
\KwOut{Query Plan $(X, EK)$}
\DontPrintSemicolon
\ForEach{$1 \leq i \leq |X|$}{ \label{line:EkRelevantStart_DP}
  $docPairs \leftarrow$ $\{(ID: j, Rank: ek_i.j) | 1 \leq j \leq k\} \cup \{(ID: null, Rank: 0)\}$\;
  $ek_i.relevant \leftarrow$ sort $docPairs$ by Rank \;
  \label{line:EkRelevantEnd_DP}
}

\tcc{$DP(i, cover)$ for cost; $EK(i, cover)$ for $EK$ values.}
\ForEach{$cover \in \mathcal{P}(gt_{sp}(q))$}{
  $DP(1, cover) \leftarrow cost_{cover}(cover, x_1)$ \;
  $EK(1, cover) \leftarrow \{ ek_1: \max_{j \in cover} ek_1.j\}$ \tcp*{dictionary}
}

\ForEach{$2 \leq i \leq |X|$}{
  \ForEach{$cover \in \mathcal{P}(gt_{sp}(q))$}{
    $DP(i, cover) \leftarrow \infty$ \;
    \ForEach{$cvr \subseteq cover$}{
      \If{$DP(i, cover) > DP(i-1, cover - cvr) + cost_{cover}(cvr, x_i)$}{
        $DP(i, cover) \leftarrow DP(i-1, cover - cvr) + cost_{cover}(cvr, x_i)$ \;
        $EK(i, cover) \leftarrow EK(i-1, cover - cvr) \cup \{ ek_i: \max_{j \in cvr} ek_i.j \}$
      }
    }
  }
}

$bestCover \leftarrow arg\,min_{|cover| \geq k' \cdot \theta_{recall}
} DP(|X|, cover)$ \;
$bestEK \leftarrow EK(|X|, bestCover)$ \;
\Return{$(X, bestEK)$}\;

\caption{Dynamic Programming}
\label{algo:RelevantDP}
\end{algorithm}

Our Dynamic Programming algorithm samples the ground truth to limit its size and enumerates its power set. Specifically, we sample $gt(q)$ several times, and each sample $gt_{sp}(q)$ has $k'$ items. When $k'$ is small, the size of its power set $\mathcal{P}(gt_{sp}(q))$ is limited to $2^{k'}$. We can then use $DP(i, cover)$ to indicate the cost of covering a item subset $cover \subseteq gt_{sp}(q)$ using the first $i$ indexes in $X$. Formally, 

\begin{sloppypar}
$DP(i, cover) = $
\begin{align*}
\small
\left\{
\begin{aligned}
cost_{cover}(cover, x_1) &\quad \text{when}~i=1, \\
\min_{cvr \subseteq cover} \Big( DP(i-1, cover - cvr) \Big. & \\
\Big. + cost_{cover}(cvr, x_i)\Big) &\quad \text{when}~2\leq i \leq |X|.
\end{aligned}
\right.
\end{align*}
where $cost_{cover}(cvr, x_i) = \max_{j \in cvr} cost_{idx}(q,x_i,ek_i.j)$ is the cost to cover item subset $cvr$ using index $x_i$. The space complexity is $O(2^{k'} \cdot |X|)$ for keeping only $DP(i-1, cover), DP(i, cover), EK(i-1, cover)$, and $EK(i, cover)$. The time complexity of DP is $O(|X|\cdot 2^{2k'})$, bringing the overall time complexity down to $O(k' \cdot \sum_{x_i \in X} x_i.dim + |X| k' \log k' + |X|\cdot 2^{2k'})$. In practice, we set $k'$ to a small constant, say $5$, to get acceptable accuracy, and the overall time complexity becomes $O(\sum_{x_i \in X} x_i.dim + |X|)$, which is linear in the sum of dimensions and number of indexes.
Algorithm~\ref{algo:RelevantDP} is the pseudo code for a sample $gt_{sp}(q)$. 
\end{sloppypar}

In \sys, we use both algorithms in different conditions. When the number of indexes $|X|$ is no more than $3$, we apply the search algorithm. Otherwise, we apply Algorithm~\ref{algo:RelevantDP} for better efficiency.

\subsubsection{Cost Estimator and Recall Estimator}
\label{section:RecallAndCostEstimator}
~

Here we explain how to estimate cost and recall for the proposed algorithms above. The idea is to use a small sample of training data and historical queries so that we can afford multiple index creations and index scans for cost estimation, as well as full scans for ground truth and recall estimation.

\begin{figure}[t]
        \centering
        \includegraphics[width=.88\linewidth]{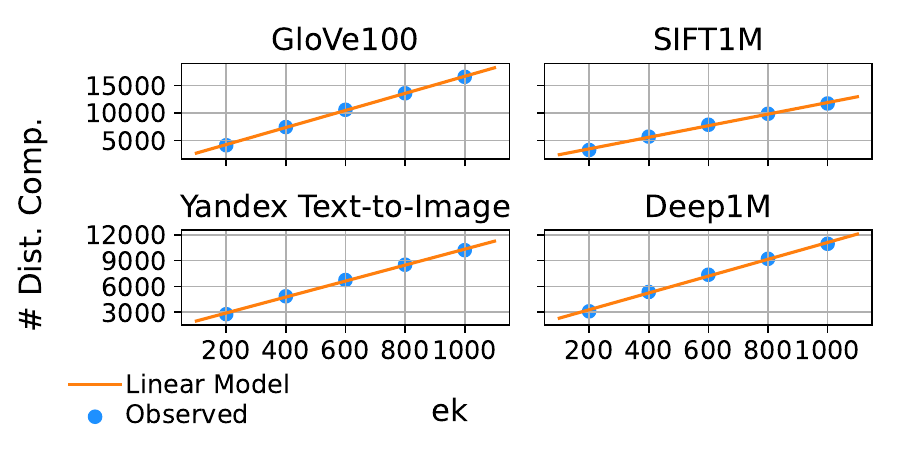}
        \vspace{-3mm}
    \caption{Observed $numDist(q,x_i,ek_i)$ using HNSW on four example datasets can be represented by linear models. Similar observation exists for DiskANN~\cite{Zhu:Arxiv2025:Mint}.}
    \label{fig:NumDistEstimate}
\end{figure}

\begin{figure}[t]
        \centering
        \includegraphics[width=.88\linewidth]{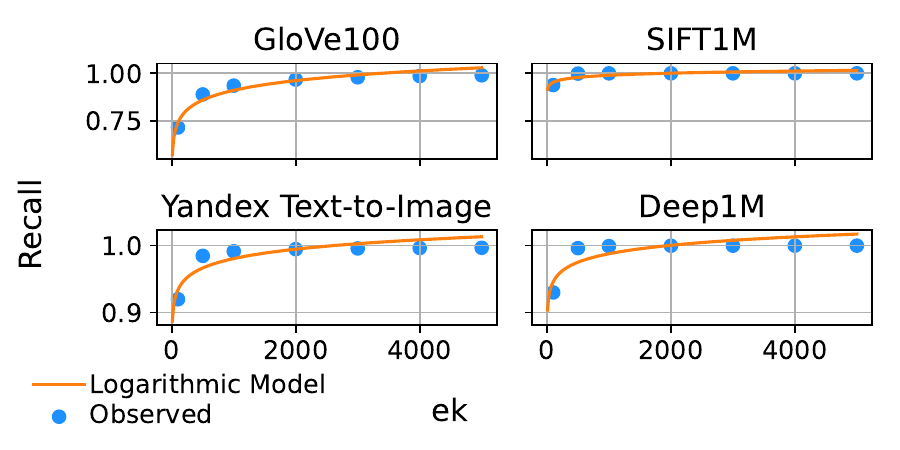}
        \vspace{-3mm}
    \caption{Observed recall using HNSW on four example datasets can be represented by logarithmic models. Similar observation exists for DiskANN~\cite{Zhu:Arxiv2025:Mint}.}
    \label{fig:RecallEstimate}
\end{figure}

\begin{sloppypar}
In practice, we use training data to estimate the cost of scanning an index. As defined in Equation~\ref{equation:CostIndexScan}, $cost_{idx}(q, x_i, ek_i) = x_i.dim \cdot numDist(q,x_i,ek_i)$. $x_i.dim$ is the sum of dimensions of underlying columns. Then the key is to estimate $numDist(q,x_i,ek_i)$, which is the number of distance/score computations. To our knowledge, graph-based vector indexes like HNSW and DiskANN have no closed-form solution for $numDist(q,x_i,ek_i)$. Therefore, we develop a simplified cost estimator. As observed in Figure~\ref{fig:NumDistEstimate}, $numDist(q,x_i,ek_i)$ is almost linear in $ek_i$ when $ek_i \geq 100$ for HNSW indexes on four different datasets. So we sample the database and queries to get training data. Then we train a linear model for each single-column index in the database. For multi-column indexes, the number of column combinations is large, and we do not train a model for each combination. Instead, we heuristically use the average slopes and intercepts across columns. We have similar observations for DiskANN and train a linear model for each column too. 
\end{sloppypar}

We use the same training data for the recall estimator as for the cost estimator. Because training data is relatively small, we can perform a full scan for each training query to find the ground truth nearest neighbors. 
As observed in Figure~\ref{fig:RecallEstimate}, we use a logarithmic function of $ek_i$ to fit recall for both HNSW and DiskANN. 

As we will see in Section~\ref{section:Experiments}, the cost and recall estimators perform reasonably well in practice.
The training data size is only 1\% of the database and the training time is between 3 and 32 seconds.
Note that our algorithms are flexible and can take any cost and recall estimators (Figure~\ref{fig:overview}) if more accurate estimators become available.


\section{Configuration Searcher}
\label{section:ConfigurationSearcher}


Configuration Searcher is the main module that searches for the best configuration to solve Problem~\ref{problem:MultiVectorIndexTuning}. It explores various configurations, and for each promising configuration, it asks the Query Planner ``if these indexes are created, what estimated cost and recall will the best query plan have'' (similar to a what-if call in a relational database~\cite{Surajit:SIGMOD1998:WhatIf}).
Next, we prove its NP-Hardness and propose a beam search algorithm.

\subsection{NP-Hardness of Configuration Search}
\label{section:ConfigurationSearchNP}

We prove the NP-Hardness of Configuration Search by reducing from the Densest $\mathtt{k}$-Subgraph problem~\cite{Chaudhuri:TKDE2004:IndexSelection, Feige:TechReport1997:DensestKSubgraph}. We replace $\mathtt{k}$ using $\mathtt{g}$ hereafter to distinguish $\mathtt{k}$ from the top-$k$ for queries.

\begin{problem}[Densest $\mathtt{g}$-Subgraph]
Given a graph $\mathtt{G}=(\mathtt{V}, \mathtt{E})$, find a subgraph $\mathtt{G}_{sub}=(\mathtt{V}_{sub}, \mathtt{E}_{sub})$ with exactly $\mathtt{g}$ vertices such that the number of edges in the subgraph is maximal.
\end{problem}

\begin{thm}
The Configuration Search problem (i.e, Problem~\ref{problem:MultiVectorIndexTuning}: Multi-Vector Search Index Tuning) is NP-Hard.
\end{thm}

The idea is to map each vertex to an index and each edge to a cost benefit of a query. When two vertices are in $\mathtt{G}_{sub}$, the query can benefit from the corresponding indexes.

\begin{proof}(\textbf{Sketch})
Let each vertex $\mathtt{v}^i \in \mathtt{V}$ be a single-column index $x^i$, each edge $(\mathtt{v}^i, \mathtt{v}^j) \in \mathtt{E}$ be a two-column query $q^{i,j}$ in Problem~\ref{problem:MultiVectorIndexTuning}, each probability $p^{i,j}$ be the same constant, and each query find the same top-$k$.
Then let $\mathtt{g}$ be $\theta_{storage}$, and $\theta_{recall} = 100\%$.
Finally, let
\begin{align*}
latency(q^{i,j}, X) =
\left\{
\begin{aligned} 
\small
k  &\quad \text{when}~\mathtt{v}^i \in \mathtt{G}_{sub} \textbf{~and~} \mathtt{v}^j \in \mathtt{G}_{sub}, \\
\small
\mathtt{L} &\quad \text{otherwise}.
\end{aligned}
\right. 
\end{align*}
where $\mathtt{L} \gg k$ is a long latency of scanning almost the whole database.
This latency function is possible in practice. 
Specifically, let each $q^{i,j}$ be a vector of all 0 but the $i$-th and $j$-th dimensions are 1. Let ground truth $gt(q^{i,j})=gt^i \cup gt^j$, where $|gt^i|=|gt^j|=k/2$. In $gt^i$, the vectors' $i$-th dimension is 1 and other dimensions are 0. Similarly, $gt^j$ has vectors with only $j$-th dimension equal to 1.
All other dummy vectors are $[0.1, ..., 0.1]$.
So ground truth vectors have $score=1$ and dummy vectors have $score=0.2$.
Plus, $gt^i$ rank at the top of  $x^i$ but $gt^j$ rank near the bottom of $x^i$ (by $score$ on $\textbf{v}_i$), and $gt^j$ rank at the top of $x^j$ but $gt^i$ rank near the bottom of $x^j$ (by $score$ on $\textbf{v}_j$). 
As a result, when $\mathtt{v}^i, \mathtt{v}^j \in \mathtt{G}_{sub}$, meaning $x^i$ and $x^j$ are both picked, the latency is $k$ by retrieving $gt^i$ from $x^i$ and $gt^j$ from $x^j$. Otherwise, we must scan other dummy items in latency $\mathtt{L}$.


Now each subgraph $\mathtt{G}_{sub}$ can be mapped to a configuration $X$ and vice versa. 
The total latency is $(|\mathtt{E}| - |\mathtt{E}_{sub}|) \cdot \mathtt{L} + |\mathtt{E}_{sub}| \cdot k = |\mathtt{E}| \mathtt{L} -|\mathtt{E}_{sub}| (\mathtt{L} - k)$.
So maximizing $|\mathtt{E}_{sub}|$ is equivalent to minimizing the total latency and vice versa.

\end{proof}

\subsection{Beam Search Algorithm}
\label{section:ConfigurationSearchAlgorithm}

\begin{algorithm}[t]
\small
\KwIn{Query workload $W$; Constraints $\theta_{storage}$ and $\theta_{recall}$; Subset difference $di$; Seed limit $se$; Beam width $b$; Improvement threshold $im$; Query Planner.}
\KwOut{Configuration $X^*$}
\DontPrintSemicolon

$Cand(q_i) \leftarrow \{ x ~|~ x.vid \subseteq q_i.vid \text{~and~} |x.vid| \geq |q_i.vid| - di \}$ \;
\label{line:BeamSearchQueryCandidate}

$Candidate \leftarrow \bigcup_{q_i} Cand(q_i)$ \;
\label{line:BeamSearchAllCandidate}

$Seed \leftarrow \bigcup_{q_i} \{X ~|~ X \subseteq Cand(q_i) \text{~and~} |X| \leq se \allowbreak \}$ \;
\label{line:BeamSearchSeed}

$Best \leftarrow$ $b$ lowest-cost configurations in $Seed$ that satisfy constraints $\theta_{storage}$ and $\theta_{recall}$ \;
\label{line:BeamSearchInitial}

\Repeat{\text{Cost improvement} $\leq im$}{
\label{line:BeamSearchLoopStart}
  $Config \leftarrow \varnothing$ \;
  \ForEach{$X \in Best, x \in Candidate$}{
      $X' \leftarrow X \cup \{x\}$ \;
      Get plans for all queries using $X'$ from Query Planner and remove unused indexes from $X'$ \;
      \label{line:BeamSearchCallQueryPlanner}
      
      \If{$X'$ satisfies constraints $\theta_{storage}$ and $\theta_{recall}$}{
        $Config \leftarrow Config \cup X'$ \;
      }
  }
  $Best \leftarrow$ $b$ lowest-cost configurations in $Config$ \;
}
\label{line:BeamSearchLoopEnd}

\Return{lowest-cost configuration $X^*$ that we have ever seen}

\caption{Beam Search}
\label{algo:BeamSearch}
\end{algorithm}

We use a beam search algorithm to find optimal configurations as shown in Algorithm~\ref{algo:BeamSearch}. We begin by finding a set of candidate indexes from all queries (Line~\ref{line:BeamSearchQueryCandidate} and \ref{line:BeamSearchAllCandidate}). For each query $q_i$, the candidate is a subset whose size is close to $q_i.vid$. For instance, when $|q_i.vid| = 5$ and $di=2$, candidates of $q_i$ are subsets with $\geq 3$ columns. Then we create a set of seed configurations from all queries (Line~\ref{line:BeamSearchSeed}). For each query $q_i$, the seed configurations are all subsets of $Cand(q_i)$ with no more than $se$ indexes.
Then we keep the best $b$ configurations in Line~\ref{line:BeamSearchInitial}, and perform beam search from Line~\ref{line:BeamSearchLoopStart} to Line~\ref{line:BeamSearchLoopEnd}. Within each iteration, we try adding one more index to the existing configuration without violating constraints. The loop stops when the cost improvement $\leq im$.

Default parameters are configured empirically. For instance, large $di$ allows a few-column index for many-column queries, which encourages sharing the index among queries. 
An overly large $di$ brings too many candidates, which slows down the search. So we set $di=2$. We set $se=2$ to ensure a big enough pool of configurations is checked. 
Setting $im=5\%$ avoids very small improvements which can be noise.

We implement several optimizations during search. For example, we may search more than $b$ seeds at the beginning to find better configurations if $|Candidate|$ is limited.
We also cache the query plans of $(q, X)$ pairs. Instead of calling Query Planner for each $(q, X')$ in Line~\ref{line:BeamSearchCallQueryPlanner}, we first identify useful indexes $X'_q= \{x ~|~ x \in X' \text{~and~} x.vid \subseteq q.vid \text{~and~} |x.vid| \geq |q.vid| - di\}$. Then, if the query plan of $(q, X'_q)$ is cached, we use the cached plan without calling Query Planner. Similarly, to save running time, we cache and pass relevant $ek$ to Query Planner if $ek$ has been computed for a $(q, x)$ pair where $x$ is an index.

\section{Experiments}
\label{section:Experiments}

We answer the following questions:
\textbf{Q1:} What is the quality of indexes and plans generated by \sys for various workloads, compared with state-of-the-art solutions? (Section~\ref{sec:exp:end2end:fixed}).
\textbf{Q2:} What is the scalability of \sys? How is the runtime of \sys affected by key factors such as query number and storage constraint? (Section~\ref{sec:exp:scalability}).

\subsection{Setting}
We implement the Configuration Searcher and Query Planner in Python, and the underlying indexes in C++. All experiments are conducted on a single machine with an AMD EPYC 7763 64-core processor, 64GB of DDR4 memory, and running Ubuntu 24.04.1 LTS. We repeat each experiment with different random seeds five times and report the mean values.\footnote{\revise{The code is not open-sourced due to the company policy, but the detailed Algorithms~\ref{algo:RelevantDP} and~\ref{algo:BeamSearch} are sufficient to reproduce all results of the paper. In addition, we release the datasets and workloads in~\cite{MintRepo}.}}

\begin{table}[] 
\centering
\footnotesize
\begin{tabular}{@{}llcc@{}}
\toprule
    \textbf{Dataset} & Type & \# rows & \# dimensions \\
\toprule
    GloVe25 & \multirow{8}{1cm}{Semi-Synthetic}  & 1M & 25\\
    GloVe50 & & 1M & 50\\
    GloVe100 & & 1M & 100\\
    GloVe200 & & 1M & 200\\
    SIFT1M & & 1M & 128 \\
    Deep1M & & 1M & 96 \\
    Music & & 1M & 100 \\
    Yandex Text-to-Image & & 1M sample & 200 \\
\hline
    NewsImage & \multirow{9}{*}{Real} & 0.1M & 512 \\
    NewsTitle & & 0.1M & 512 \\
    NewsDescription & & 0.1M & 768 \\
    NewsContent & & 0.1M & 768 \\
    AmazonTitle & & 10M & 768 \\
    AmazonImage & & 10M & 512 \\
    AmazonDescription & & 10M & 768 \\
    AmazonFeature & & 10M & 768 \\
    AmazonDetail & & 10M & 768 \\
\bottomrule
\end{tabular}
\caption{Datasets, each of which is a column (i.e., feature).}
\label{tbl:columns}
\end{table}

\begin{table}[] \centering
\footnotesize
\begin{tabular}{@{}llcccc@{}}
\toprule
    \multirow{2}{*}{\textbf{Workload}} & \multirow{2}{1cm}{Dataset Type} & \multirow{2}{*}{\# rows} & \multirow{2}{*}{\# cols} & \multirow{2}{*}{\# queries} & \multirow{2}{1.5cm}{Avg. \# cols per query} \\
    & & & & & \\
    \toprule
    \SynNaive & \multirow{3}{1cm}{Semi-Synthetic} & 1M & 3 & 4 & 2.00\\
    \SynLargeSimple & & 1M & 8 & 12 & 2.25 \\
    \SynLargeComplex & & 1M & 8 & 12 & 4.08\\
\hline    
    \News & \multirow{2}{1cm}{Real} & 0.1M & 4 & 6 & 2.50 \\
    \Amazon & & 10M & 5 & 8 & 3.25 \\
\hline
\end{tabular}
\caption{Workloads.}
\label{tbl:datasets}
\end{table}

\subsubsection*{\textbf{Datasets and Workloads}}
To evaluate the performance of \sys, we use a mix of semi-synthetic datasets and real-world datasets~(Table~\ref{tbl:columns}), \revise{and five workloads on these datasets (Table~\ref{tbl:datasets}). In each workload, we draw query weights i.i.d. from Uniform(0,1) and normalize them to sum to one.}

We use semi-synthetic datasets because most real-world multi-modal datasets have only two columns (text and image), in which case our recommended indexes are the same as the \baseline baseline. To experiment with more than two columns, we generate semi-synthetic datasets by combining single-feature embedding datasets including GloVe25~\cite{Pennington:EMNLP2014:GloVeDataset}, GloVe50, GloVe100, GloVe200, SIFT1M~\cite{Jegou:TPAMI2011:SiftDataset}, Yandex Text-to-Image~\cite{Simhadri:Neurips2021:BigANN-Benchmark}, Deep1M~\cite{Yandex:CVPR2016:DeepDataset}, and Music~\cite{Morozov:2018:MusicDataset}.
We sample 1M rows from each single-feature dataset and treat each as a separate column, mimicking the multi-feature scenario. \revise{
Using these features, we create three workloads: \SynNaive, \SynLargeSimple, and \SynLargeComplex.
\SynNaive contains three columns (GloVe100, SIFT1M, and Yandex Text-to-Image), 1M rows, and four manually crafted queries for the case study.
\SynLargeSimple and \SynLargeComplex contain all eight semi-synthetic columns in Table~\ref{tbl:columns} and 1M rows. Each workload has 12 queries. For each query, each column is independently selected with probability $\mathtt{p}$. Intuitively, a smaller $\mathtt{p}$ generates simpler queries with fewer columns. 
We use $\mathtt{p}=0.3$ for \SynLargeSimple and $\mathtt{p}=0.5$ for \SynLargeComplex. For example, a query in \SynLargeSimple is on columns GloVe200 and Music, and a query in \SynLargeComplex is on GloVe100, SIFT1M, and Deep1M.}

One real-world dataset is \News, consisting of 0.1M rows and four embeddings extracted from titles, descriptions, contents, and images of BBC News~\cite{li2024latesteval}.
We get the embeddings for images and titles using the CLIP model~\cite{radford2021learning}, which handles images and short text with less than $77$ tokens.
For longer text, such as news description and content, we use Nomic-BERT model~\cite{nussbaum2024nomic}. 
\revise{The \News workload has six randomly-generated queries on the four real embeddings. 
In each query, column are selected independently with probability $\mathtt{p}=0.5$. One example query is on columns NewsTitle and NewsContent.}

We also experiment with a large-scale \Amazon~\cite{Hou:2024:Bridging} dataset including 10M rows and five embeddings to verify the performance gain in Section~\ref{sec:exp:scalability}. \revise{
Its image embeddings are generated using CLIP, and its title, description, feature, and detail embeddings are generated using Nomic-BERT. It has eight queries. For each query, each column is independently selected with probability $\texttt{p}=0.5$. One example query is on columns AmazonTitle and AmazonFeature.}

\begin{figure}[t]
    \centering
    \includegraphics[width=.93\linewidth]{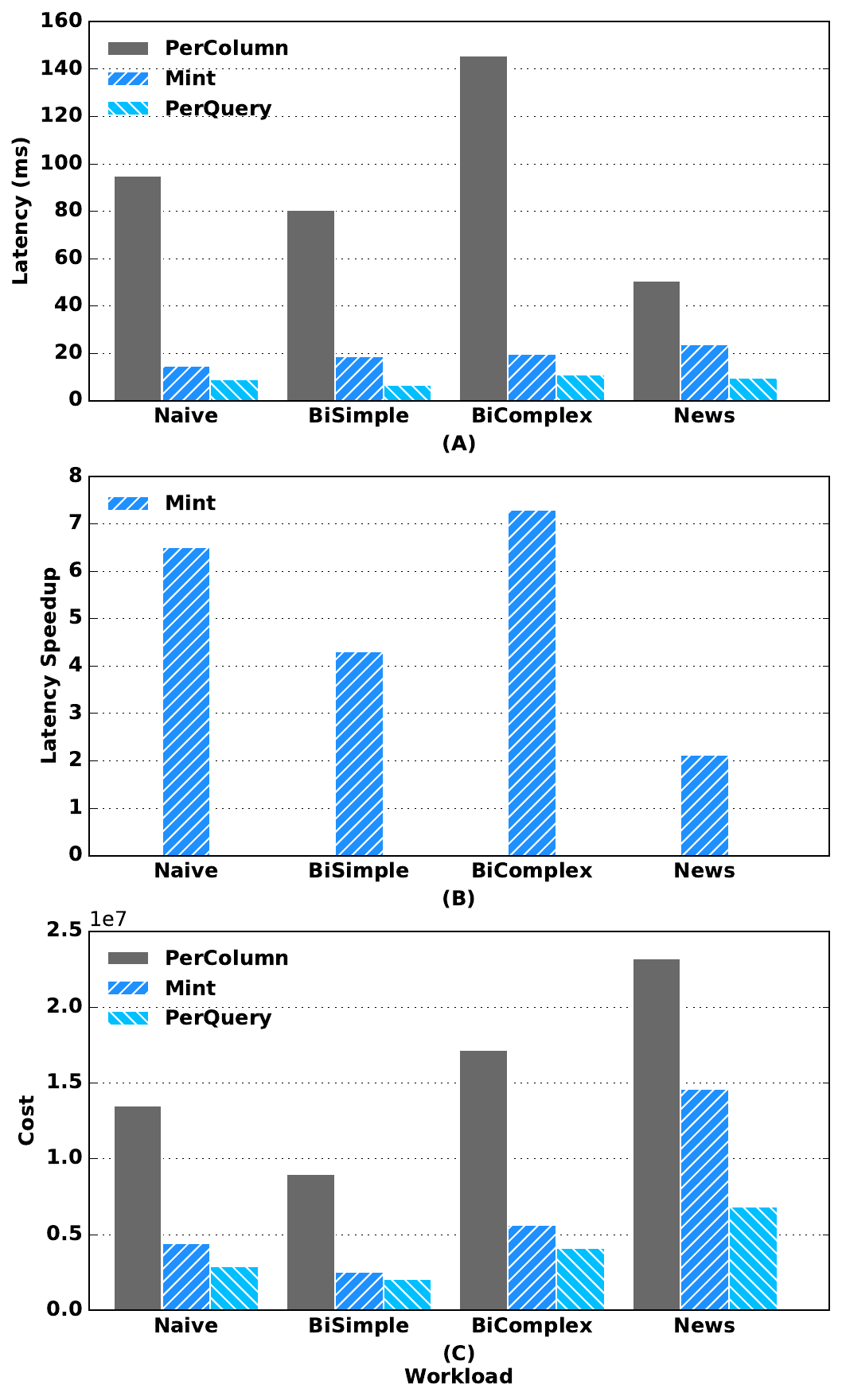}
    \caption{Using HNSW, (A) \sys outperforms \baseline on all four workloads, and \sys's latency is close to the optimal \PerQuery's, although \PerQuery uses 33\% to 50\% more storage. (B) The speedup ranges from $2.1\times$ to $7.2\times$. (C) \sys outperforms \baseline because \sys has lower cost.}
    \label{fig:ExpEndToEnd_HNSW}
\end{figure}

\subsubsection*{\textbf{Approaches}}
\begin{sloppypar}
We compare \sys against two baselines: \baseline and \PerQuery.
\baseline means one-index-per-column~\cite{wang:SIGMOD2021:milvus,zhang:OSDI2023:vbase}. 
It constructs an independent index for each column in the database. Given a workload, each query is answered by the indexes corresponding to the columns involved in the query. 
\PerQuery means one-index-per-query. It creates one index for each query in the workload, so it violates the storage constraints in all workloads we experiment with. However, we still consider it to be a lower bound of latency.
All approaches (\sys, \baseline, and \PerQuery) use two widely adopted ANN indexes, HNSW and DiskANN.
In \sys, we sample 1\% of the data to train the cost estimator and recall estimator (Section~\ref{section:RecallAndCostEstimator}).
\end{sloppypar}

\subsubsection*{\textbf{Metrics and Constraints}}
We use cosine similarity as the distance measure in all experiments.
We measure the \emph{latency} and the cost as a proxy for latency (Section~\ref{sec:DefQueryPlanning}).  
Given a workload of queries with probabilities, we measure the probability-weighted average latency and cost across queries.

We measure \emph{recall@100} for the recall constraint in Problem~\ref{problem:MultiVectorIndexTuning}. 
We set the recall constraint to $90\%$ for semi-synthetic workloads with more items, and $97\%$ for real workloads with many fewer items to make the task non-trivial.

The storage in bytes is proportional to the number of indexes when a given workload has a fixed number of items, and our maximum degree of HNSW and DiskANN defaults to $16$. So we show the number of indexes as the storage. (Recall in Section~\ref{section:FrameworkOverview} that the index graph size is independent of the vectors' dimensions.)

The default storage constraint is the number of columns because \baseline requires one index per column. As Table~\ref{tbl:datasets} shows, the storage constraints of \SynNaive, \SynLargeSimple, \SynLargeComplex, and \News are $3, 8, 8$, and $4$, respectively. We apply the default storage to \baseline and \sys for a fair comparison. \PerQuery requires one index per query and uses $4, 12, 12$, and $6$ storage, which violates the constraint but is shown as a reference. In addition, since \sys is designed to be flexible with storage, we evaluate \sys under various storage constraints.

\subsection{End-to-end Performance}\label{sec:exp:end2end}

Here we evaluate the end-to-end performance of \sys, \baseline, and \PerQuery across \SynNaive, \SynLargeSimple, \SynLargeComplex, and \News workloads. 

\subsubsection*{\textbf{Fixed Storage}}\label{sec:exp:end2end:fixed}
\begin{figure}
    \centering
    \includegraphics[width=.93\linewidth]{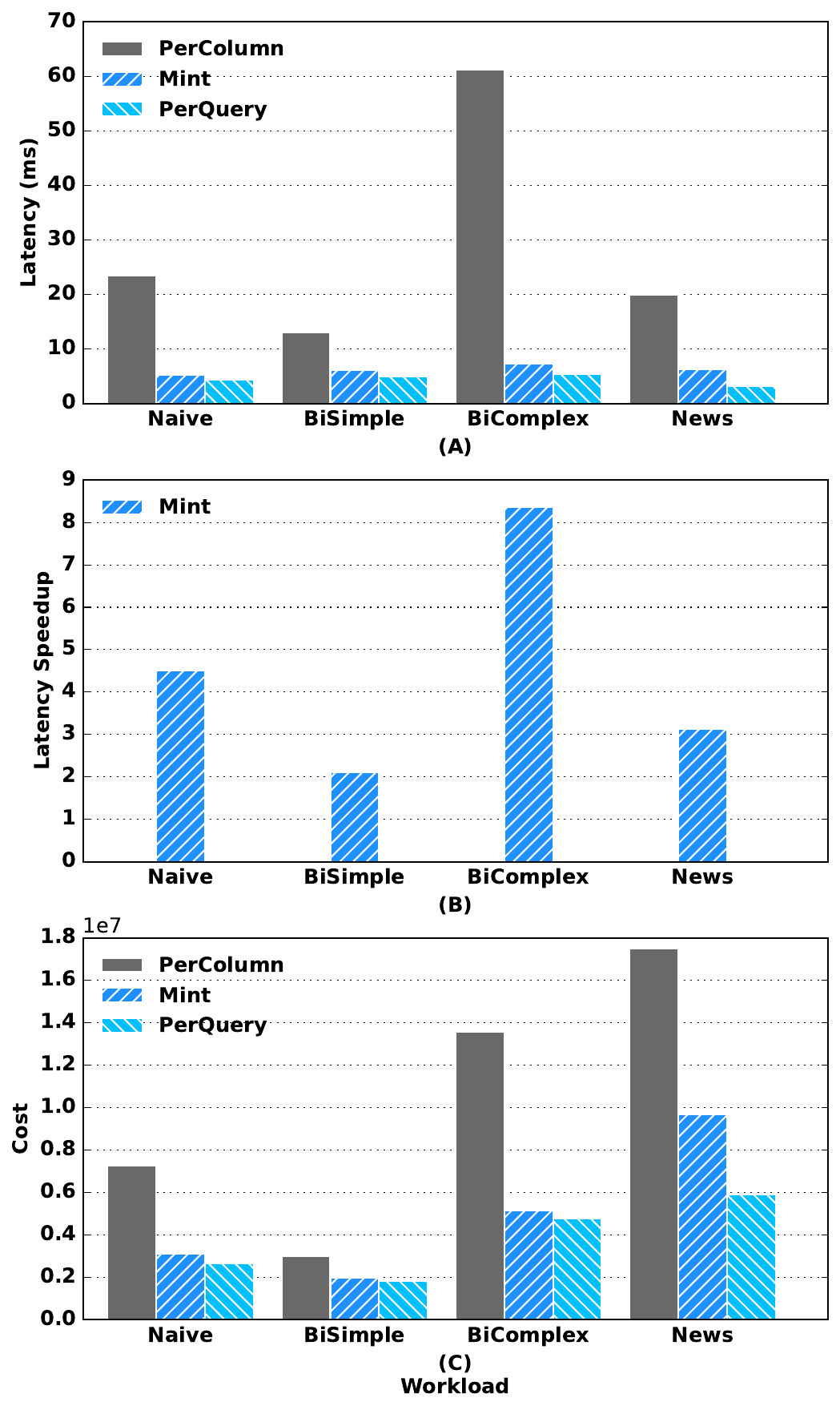}
    \caption{Using DiskANN, (A) \sys outperforms \baseline on all four workloads, and \sys's latency is close to the optimal \PerQuery's, although \PerQuery uses 33\% to 50\% more storage. (B) The speedup ranges from $2.1\times$ to $8.3\times$. (C) \sys outperforms \baseline because \sys has lower cost.}
    \label{fig:ExpEndToEnd_DiskANN}
\end{figure}

\begin{figure}
    \centering
    \includegraphics[width=.9\linewidth]{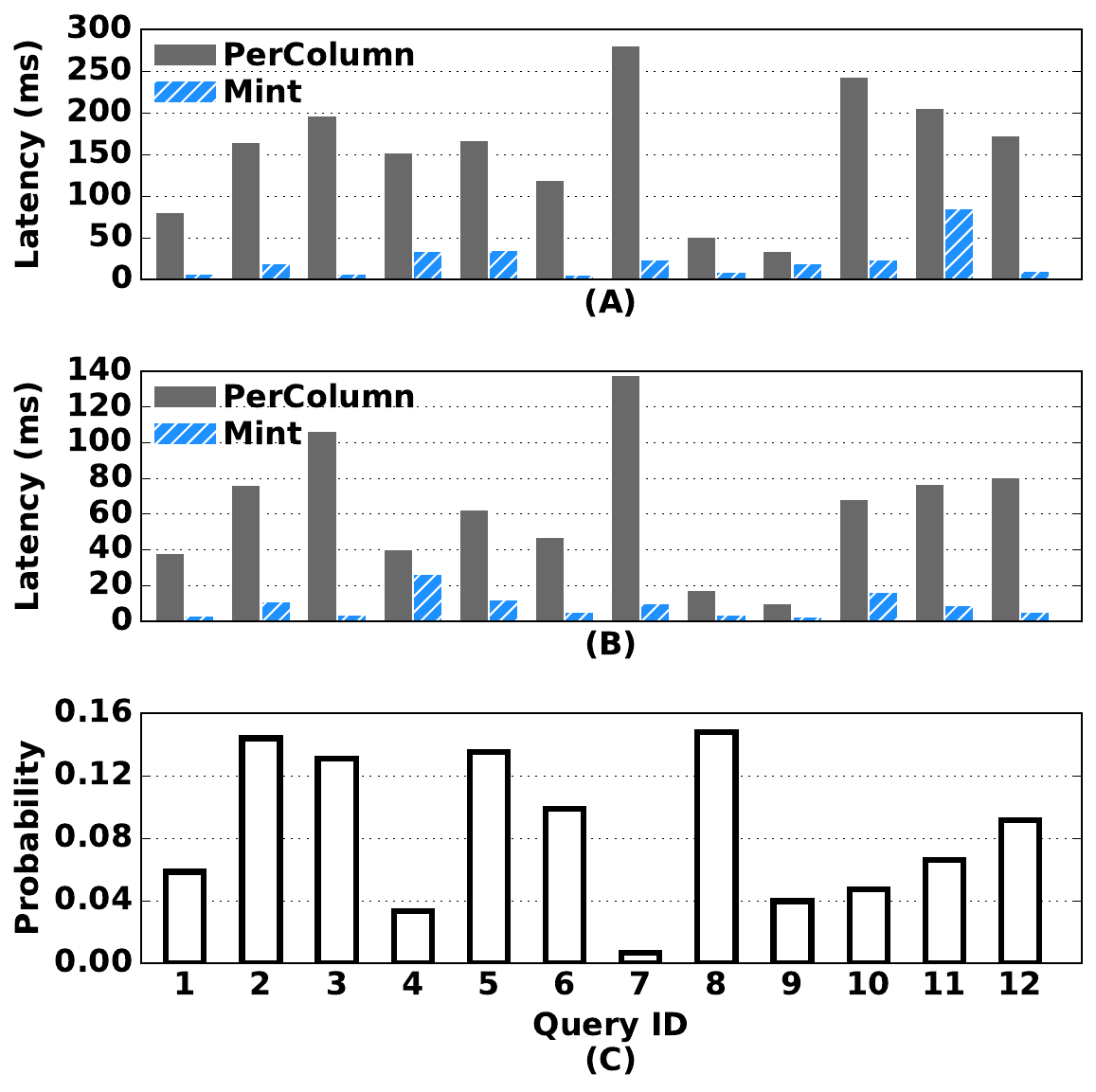}
    \caption{Query latency for (A) HNSW and (B) DiskANN in \SynLargeComplex workload and (C) queries probabilities.}
    \label{fig:ExpEndToEnd_Breakdown_SynLargeComplex}
\end{figure}

The end-to-end latency, speedup, and cost are in Figure~\ref{fig:ExpEndToEnd_HNSW} (using HNSW) and Figure~\ref{fig:ExpEndToEnd_DiskANN} (using DiskANN). \sys and \baseline use a fixed storage budget that equals the number of columns. \PerQuery, as a reference, uses additional storage (33\% more on \SynNaive and 50\% more on \SynLargeSimple, \SynLargeComplex, and \News).
Across all workloads and indexes, 
\sys consistently outperforms \baseline in terms of latency. The speedup ranges from $2.1\times$ to $7.2\times$ for HNSW indexes and $2.1\times$ to $8.3\times$ for DiskANN indexes.
In addition, \sys achieves comparable latency compared to the latency lower bound obtained from \PerQuery in almost all scenarios. \sys outperforms \baseline because \sys has a lower cost in each workload. 
Due to space limitations, we show each query's latency in the \SynLargeComplex workload as an example in
Figure~\ref{fig:ExpEndToEnd_Breakdown_SynLargeComplex}. \sys outperforms \baseline on all queries using HNSW and DiskANN. The speedup ranges from $1.5\times$ to $27.8\times$ with median=$11.1\times$. 

The performance gains of \sys stem from its use of multi-column indexes. 
For example, in Figure~\ref{fig:ExpEndToEnd_Breakdown_SynLargeComplex}B, \baseline has much higher latency than \sys, especially on queries with more columns, such as Query 7 (6 columns, 138 ms), Query 2 (4 columns, 76 ms), and Query 3 (3 columns, 106 ms). In comparison, \sys
uses 4-column and 3-column indexes to answer these three queries, achieving latency speedups from $6.8\times$ to $27.8\times$. In addition, these multi-column indexes are shared by multiple queries, which helps \sys meet the storage constraint.

\subsubsection*{\textbf{Varying Storage}}\label{sec:exp:adaptation}
\begin{figure}
    \centering
    \includegraphics[width=.9\linewidth]{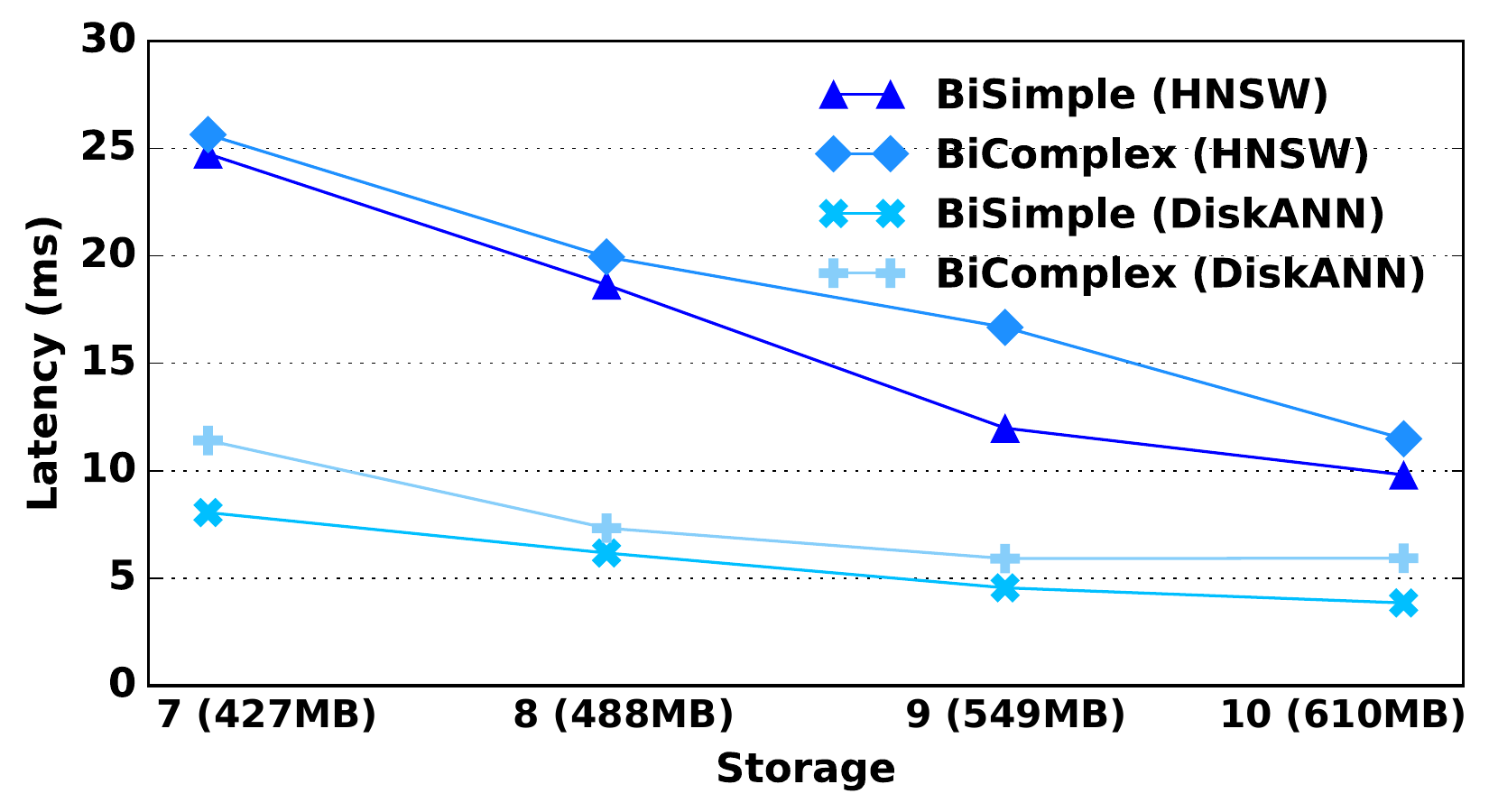}
    \caption{\sys finds configurations with lower latency when storage increases. 
    }
    \label{fig:ExpStorageLatencyAll}
\end{figure}

To explore how varying storage constraints affects the effectiveness of \sys, we ran \sys on \SynLargeSimple and \SynLargeComplex with the number of indexes ranging from 7 to 10.
As shown in \Cref{fig:ExpStorageLatencyAll}, for all workloads and ANN indexes, \sys achieves lower latency when more indexes are created.
So \sys works under different storage constraints.


\subsubsection*{\textbf{Case Study on \SynNaive}}

\begin{figure}
    \centering
    \includegraphics[width=.9\linewidth]{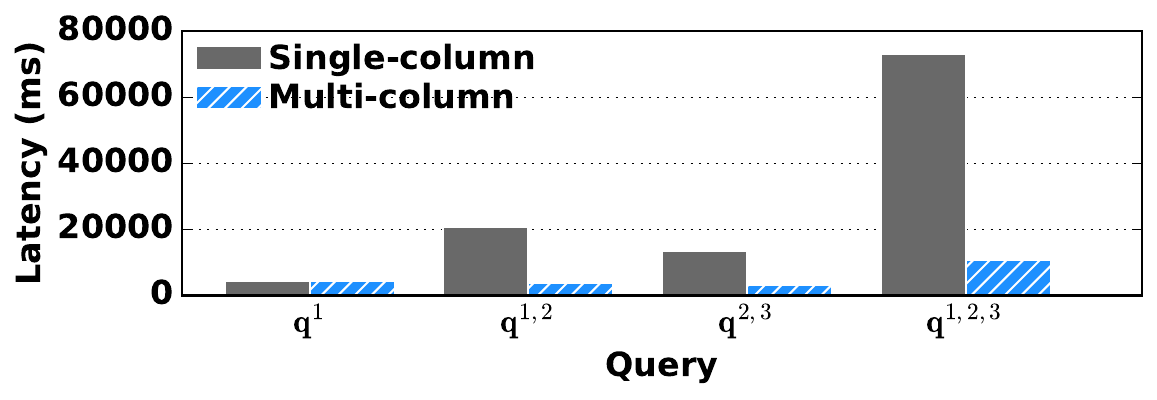}
    \caption{Query latency in \SynNaive workload using DiskANN.}
    \label{fig:ExpEndToEnd_CaseStudy_Naive}
\end{figure}

We present a case study to show how \sys benefits from multi-column plans. 
Table~\ref{tab:case_study:query_plans} shows the query plans for the \SynNaive workload using DiskANN. 
The columns are numbered $1, 2$, and $3$. Each plan shows the indexes followed by the numbers of extracted items (i.e., $ek$). Each query's latency is in Figure~\ref{fig:ExpEndToEnd_CaseStudy_Naive}.

As the number of columns in a query grows, the query benefits more from multi-column indexes. 
For example, starting from $q^{1,2}$, the multi-column plan extracts fewer items than the single-column plan ($1625$ vs. $3106$). 
Also, as $x^{1,2}$ is on the same columns as $q^{1,2}$, it avoids re-ranking in the plan of $x^1$ and $x^2$. For the three-column query $q^{1,2,3}$, although the multi-column plan scans two indexes $x^{1,2}$ and $x^{2,3}$, it extracts only 13\% of the items of the single-column plan. The reduction of extracted items leads to much lower latency.

\begin{mdframed}
\textit{Key Takeaway.} \sys consistently outperforms \baseline across all workloads and indexes under the same storage constraint. \sys is close to the optimal \PerQuery although \PerQuery uses 33\% to 50\% more storage.
\sys is also flexible with different storage constraints. The use of multi-column indexes is critical for achieving superior performance.
\end{mdframed}




\begin{table}[t] \footnotesize 
\centering
\begin{tabular}{@{}l|cccc@{}}
\toprule
 & $q^1$ & $q^{1,2}$ & $q^{2,3}$ & $q^{1,2,3}$ \\ 
\midrule
\makecell{Single-column \\ plans} &
$x^1$: 1,461 &
\begin{tabular}[r]{@{}r@{}}$x^1$: 1,684\\ $x^2$: 1,422 \\ Total: 3,106 \end{tabular} &
\begin{tabular}[r]{@{}r@{}}$x^2$: 1,206\\$x^3$: 3,083 \\ Total: 4,289 \end{tabular} &
\begin{tabular}[r]{@{}r@{}}$x^1$: 7,563	
\\$x^2$: 4,412 \\$x^3$: 6,093 \\ Total: 18,068 \end{tabular} \\[2ex]
\hline
\makecell{Multi-column \\ plans} & 
$x^1$: 1,461 &
$x^{1,2}$: 1,625 & 
$x^{2,3}$: 199 & 
\begin{tabular}[r]{@{}r@{}}$x^{1,2}$: 1,248 \\$x^{2,3}$: 1,161 \\ ~~Total: 2,409 \end{tabular} \\ 
\bottomrule
\end{tabular}
\caption{Compared to single-column plans, multi-column plans extract much fewer items to meet recall constraint in \SynNaive workload.}
\label{tab:case_study:query_plans}
\end{table}


\subsection{Scalability Analysis}\label{sec:exp:scalability}
Here we show how the runtime of \sys is affected by the storage constraint and number of queries, and show that \sys still outperforms \baseline on a large-scale dataset.

\subsubsection*{\textbf{Overall runtime analysis}}
\begin{figure}
    \centering
    \includegraphics[width=.93\linewidth]{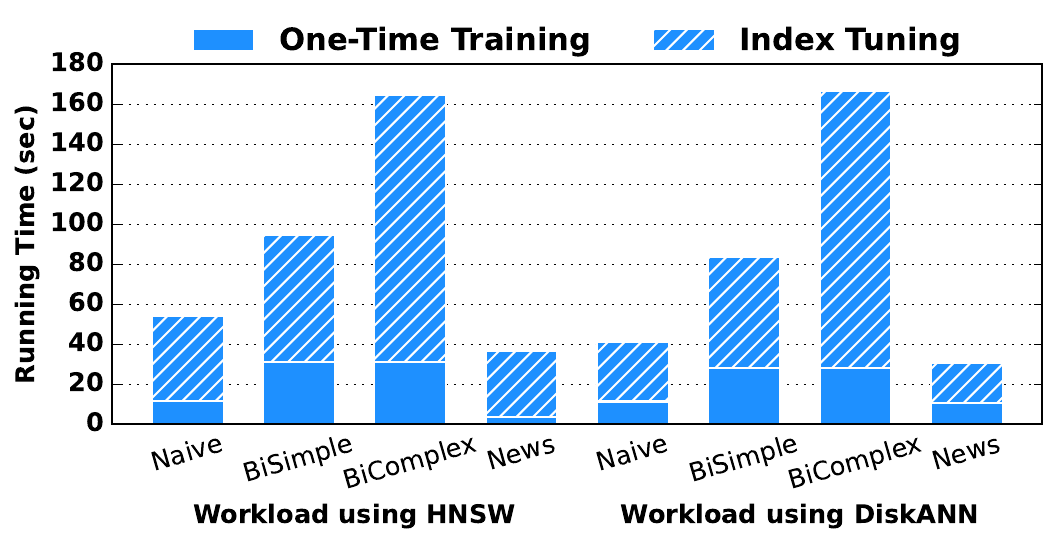}
    \caption{\sys finds configurations in minutes. The running time breaks down into one-time training and index tuning. The one-time training runs only once given a database, and its time can be amortized across different workloads.}
    \label{fig:ExpEndToEnd_RunningTime}
    \vspace{-5mm}
\end{figure}

As reported in \Cref{fig:ExpEndToEnd_RunningTime}, \sys finds configurations in minutes. The running time of \sys breaks down into one-time training, which learns models for the cost estimator and recall estimator, and index tuning, which utilizes the learned models in estimators to find indexes and plans for a given workload. 
Note that the one-time training runs only once for a given database, and its time can be amortized across different workloads.

\subsubsection*{\textbf{Storage constraint}}
\begin{figure}
    \centering
    \includegraphics[width=.88\linewidth]{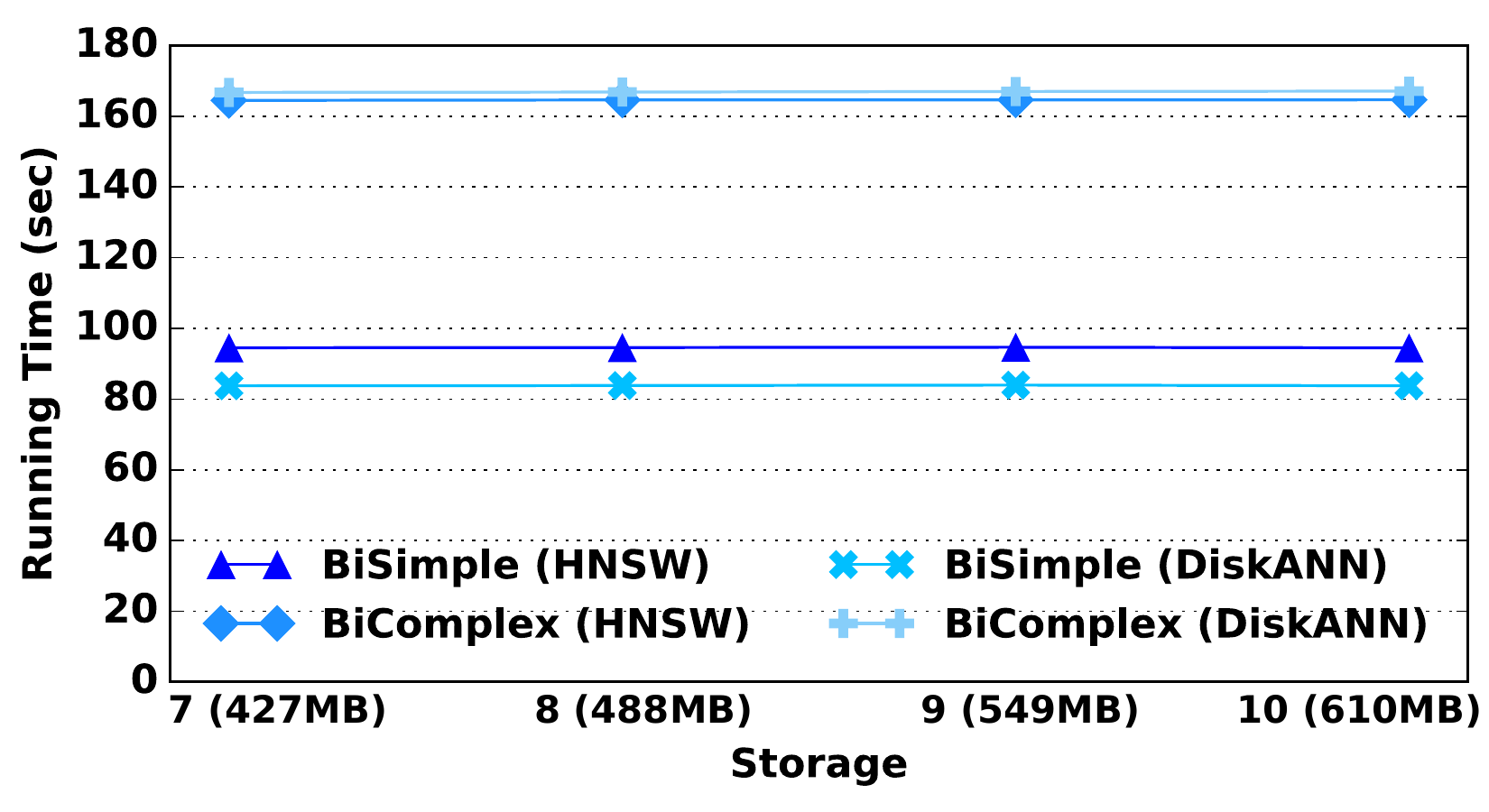}
    \caption{\sys's running time does not increase much when storage increases, because storage is within a limited range in practice.} 
    \label{fig:ExpStorageRunningTimeAll}
    \vspace{-5mm}
\end{figure}

As shown in Figure~\ref{fig:ExpStorageRunningTimeAll}, increasing the storage budget has a minimal impact on the total planning time. 
Most intermediate results are generated in the early iterations and are cached using the mechanism in Section~\ref{section:ConfigurationSearchAlgorithm}.
Thus, they can be reused for subsequent iterations. This allows more indexes to be created, increasing the storage budget without substantially increasing computation costs. 

\subsubsection*{\textbf{Query number}}
\begin{figure}
    \centering
    \includegraphics[width=.88\linewidth]{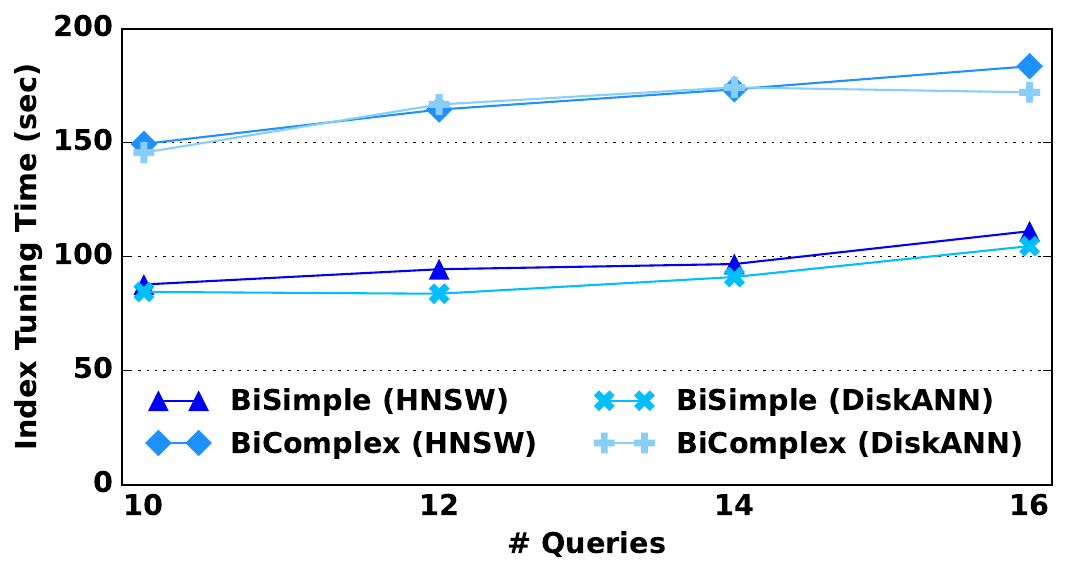}
    \caption{As workload size increases, the running time increases almost linearly.}
    \label{fig:ExpQuerySearchTimeAll}
    \vspace{-5mm}
\end{figure}

Figure~\ref{fig:ExpQuerySearchTimeAll} shows that the running time of \sys grows almost linearly with the number of queries in a workload.
This is because the runtime of Algorithm~\ref{algo:BeamSearch} is dominated by getting plans for each query given the current set of indexes (Line~\ref{line:BeamSearchCallQueryPlanner}), and the runtime of this step grows linearly with the number of queries.




\subsubsection*{\textbf{Large-scale Dataset}}
We experimented with the \Amazon~\cite{Hou:2024:Bridging} dataset with 10M rows and 5 columns using 8 queries. We observe a 5.3$\times$ query latency speed-up compared to \baseline. More details are in \cite{MintRepo}.

\begin{mdframed}
\textit{Key Takeaway.} \sys is efficient and scales well with storage constraints and workload sizes, and still outperforms \baseline on 10M rows.
\end{mdframed}





\section{Related Work}
\label{section:RelatedWork}

The single-vector search is known as Nearest Neighbor search and dates back to 1960s~\cite{Cover:TranInfTheor1967:NearestNeighbor, Bentley:CommACM1975:KdTree, Omohundro:1989:BallTree}. 
Due to the growth of vector sizes for many scenarios, such as RAG, \emph{exact} nearest neighbor search is infeasible.
Therefore, researchers have developed \emph{Approximate} Nearest Neighbor (ANN) search techniques to retrieve nearest neighbors more efficiently. 

There are several categories of ANN techniques~\cite{pan:VLDBJ2024:survey}. 
Partition-based techniques~\cite{Chen:Neurips2021:SPANN, babenko2014inverted, lv2007multi, liu2014sk, norouzi2012fast, Datar:SympCompGeometry2004:LSH, Lloyd:InfoTheory1982:KMeans, Andoni:CommACM2008:LshAnn} put similar items into the same partition. 
Tree-based techniques~\cite{muja:VISAPP2009:FLANN, Chanop:CVPR2008:PKDTree, Dasgupta:STOC2008:RPTree} split items recursively into a tree. 
Graph-based techniques organize items in graphs and are prevalent these days because of their efficiency and high recall~\cite{Aumuller:InfoSys2020:ANN-Benchmarks, Simhadri:Neurips2021:BigANN-Benchmark, Simhadri:Arxiv2024:BigANN-Benchmark, subramanya:NEURIPS2019:diskann, malkov:TPAMI2018:hnsw, Malkov:InfoSys2014:NSW, fu2017fast}. 
Quantization techniques~\cite{Gray:ASSP1984:VectorQuantization, jegou2010product, ge2013optimized, guo2020accelerating, Gao:SIGMOD2024:RabitQ, Gao:SIGMOD2025:ExtendedRabitQ} can be used by all these methods to compress vectors and improve efficiency. 
They are implemented in various industry libraries and systems~\cite{Chen:2018:SPTAG, Douze:Arxiv2024:faiss, Johnson:BigData2019:FaissGPU, Boytsov:Sisap2013:NMSLib, Yang:SIGMOD2020:PASE, guo2022manu, su2024vexless, PgVector, Pinecone, Zilliz, Lucene, AzureAiSearch}. 
All of the above ANN techniques focus on single-vector search. They are orthogonal to our Multi-Vector Search Index Tuning problem and can be plugged into our flexible framework.


In multi-vector search, items are represented by multiple vectors~\cite{pan:VLDBJ2024:survey, wang:ICDE2024:must, duong2021efficient, ding2025www}, each of which is in a separate column. The one-index-per-column approach creates an index on each column and scans multiple indexes to answer a query. For instance, Milvus~\cite{wang:SIGMOD2021:milvus} and VBase~\cite{zhang:OSDI2023:vbase} scan multiple single-column indexes in turn and merge the results to find top results. OneSparse~\cite{Chen:WWW2024:OneSparse} pushes down an intersection operation across indexes to filter out low-quality results earlier. Compared to scanning single-column indexes, we scan multi-column indexes to find the top result more quickly. In addition, we study which multi-column indexes should be created and scanned across multiple queries. 

\revise{Another line of work represents each item as a set of multiple vectors within a single semantic domain (e.g., token- or paragraph-level embeddings) and computes similarity using late-interaction functions such as MaxSim~\cite{Khattab:2020:ColBERT, Santhanam:2022:colbertv2, Scheerer:SIGIR2025:MutiVector, Bian:SIGIR2025:MultiVector}. Each item consists of a single column whose value is a variable-size set of vectors. These methods focus on fine-grained matching over a single column and do not consider multiple heterogeneous vector columns or the index tuning problem studied in this work, where queries may involve different subsets of columns.}

\revise{Some related works study multi-column vector queries but do not consider index tuning. DEG~\cite{Yin:SIGMOD2025:HybridVectorSearch} focuses on weighted queries over exactly two fixed vector columns.
HJG~\cite{Zhu:ICDE2024:MultiColumn} proposes a well-designed unified index that supports weighted queries over multiple vector columns, but it relies on a single fixed index and does not consider workload-aware index tuning. In practice, queries may have heterogeneous recall requirements and benefit from dedicated indexes on specific column subsets, which can achieve lower latency and better recall than a unified index. Our work instead studies index tuning for a query workload, selecting and configuring indexes under varying recall constraints.}

Fagin et al.'s threshold algorithms~\cite{Fagin:SIGMOD2001:FaginAlgorithms, FaginLN03} and follow-on work~\cite{BadrVodislav2015,TheobaldWS04,ChenSG11} address the problem of answering exact and approximate top-k queries.
They scan multiple sorted lists in turns until the result cannot be improved by scanning the remaining items. 
These query processing algorithms are orthogonal to our \sys algorithm, which focuses on index tuning. 
If these algorithms can provide cost and recall estimators, then they can be plugged into our framework.


The Index Tuning problem has been extensively studied for relational databases~\cite{Wu:TKDE2024:IndexTuningSurvey, chaudhuri:VLDB2007:TuningDecade, Chaudhuri:VLDB1997:AutoAdmin, Debabrata:VLDB2011:CoPhy, Bruno:SIGMOD2005:Relaxation}. 
This problem significantly differs from multi-vector search index tuning in its cost model and query planning. 
Therefore, we needed to develop new algorithms for multi-vector search index tuning.

\revise{An index can be viewed as a special case of a materialized view (MV)~\cite{Agrawal:VLDB2000:MaterializedViews, Gupta:1999:MaterializedViews}. While MVs are effective for queries involving filters, aggregations, and joins over multiple tables, our work focuses on vector indexes on a single table without such operations; thus, we do not consider MVs as a separate design choice. MVs may become relevant in extended settings such as filtered vector search~\cite{Li:VLDB2025:FilteredVectorSearch}, which we leave as future work.}

\vspace{-3mm}
\section{Conclusions}
\label{section:Conclusion}

In this paper, we propose \sys, a framework for the multi-vector search index tuning problem. In a multi-vector database, each row is an item, each column represents a feature, and each cell is a high-dimensional vector.
Given a workload of queries, \sys recommends indexes to minimize the latency while meeting the storage and recall constraints. \sys has two major modules, Query Planner and Configuration Searcher. 
We formally define the problems of Query Planning and Configuration Searching, prove they are NP-hard, and develop algorithms to solve them. 
Our experiments show a $\speedupLowerBound$ to $\speedupUpperBound$ latency speedup compared to the state of the art. 
Since \sys is designed to be flexible with estimators and ANN indexes, in future work, we plan to develop more accurate estimators for more performance gain and apply \sys to other types of indexes in addition to HNSW and DiskANN.


\section{AI-Generated Content Acknowledgment}
No section in this paper was AI-generated.


\balance
\bibliographystyle{IEEEtran}
\bibliography{ref}

\end{document}